\documentclass[superscriptaddress,aps,amsfonts,nofootinbib,notitlepage]{revtex4-2}
\usepackage{xcolor,graphicx}
\usepackage{breqn,amsmath}

\usepackage{booktabs}
\usepackage{makecell}

\usepackage{comment}

\usepackage{enumerate}

\usepackage{natbib,hyperref}

\usepackage{subcaption}

\usepackage[T1]{fontenc}

\usepackage{mathtools}

\makeatletter
\let\cat@comma@active\@empty
\makeatother

\usepackage{IEEEtrantools}

\begin{document}

\bstctlcite{IEEEexample:BSTcontrol}

\title{Healthy Horndeski cosmologies with torsion} 
\author{S. Mironov}
\email{sa.mironov\_1@physics.msu.ru}
\affiliation{Institute for Nuclear Research of the Russian Academy of Sciences, 
60th October Anniversary Prospect, 7a, 117312 Moscow, Russia}
\affiliation{Institute for Theoretical and Mathematical Physics,
MSU, 119991 Moscow, Russia}
\affiliation{NRC, "Kurchatov Institute", 123182, Moscow, Russia}

\author{M. Valencia-Villegas}
\email{mvalenciavillegas@itmp.msu.ru}

\affiliation{Institute for Theoretical and Mathematical Physics,
MSU, 119991 Moscow, Russia}

\begin{abstract}
We show that the full Horndeski theory with  both curvature and torsion can support nonsingular, stable and subluminal cosmological solutions at {\it all times}. Thus, with torsion, the usual No-Go theorem that holds in a curved spacetime is avoided. In particular,  it is essential to include the nonminimal derivative couplings of the $\mathcal{L}_{5}$ part of the Horndeski action ($G^{\mu\nu}\,\nabla_\mu \nabla_\nu \phi,$ and $(\nabla^2 \phi)^3$). Without the latter a No-Go already impedes the eternal subluminality of nonsingular, stable cosmologies.
\end{abstract}

\maketitle


\section{Introduction}

Modifications of General Relativity (GR) with scalar fields have been widely investigated mainly motivated by the need to address the  singularity issues of GR, the  Dark Energy and Dark Matter problems, and the recent phenomenological advances which promise guidelines and constraints to the vast theoretical possibilities \cite{arai2023cosmological}. 

In this work we consider the most general modification of GR with second derivatives of a scalar in the action, but with second order equations of motion, which is known as Horndeski theory \cite{horndeski1974second,nicolis2009galileon,Deffayet:2011gz,Kobayashi:2011nu,deffayet2009covariant}. It reunites under one class a wide variety of models, ranging from the cosmological constant, to k-essence and more generally, minimally coupled and non-minimally coupled scalars, such as Brans-Dicke (See for instance \cite{kobayashi2019horndeski} for a review). 

The main attractive feature of Horndeski theory -later rediscovered as Galileons \cite{nicolis2009galileon,Deffayet:2011gz,Kobayashi:2011nu,deffayet2009covariant}- is that it permits to violate without obvious pathologies the Null Energy Condition (NEC) \footnote{See for instance \cite{rubakov2014null} for a review.}, upon which the singularity theorems of Penrose and Hawking hold \cite{Penrose:1964wq, Hawking:1965mf}. Indeed it has been proven possible to build {\it locally} stable and subluminal {\it bouncing} cosmologies, which however, generally have an issue with {\it global} stability \cite{Evslin:2011vh,Easson:2011zy,Sawicki:2012pz,Rubakov:2016zah,Libanov:2016kfc, Kobayashi:2016xpl,Kolevatov:2016ppi,Cai:2017dyi,Mironov:2019fop,Akama:2017jsa,Cai:2016thi,Cai:2017tku,Creminelli:2016zwa,Kolevatov:2017voe}\footnote{See also \cite{Ijjas:2016tpn} and the  stability issues discussed in \cite{Dobre:2017pnt}.}. In other words, even if the solutions are {\it locally} healthy around the bounce - or the most physically relevant phase - an instability will certainly happen at some time earlier or later in the evolution of the Universe. The certainty of instabilities at some time is what we call {\it global} stability issues, which were established as No-Go theorems in \cite{Libanov:2016kfc, Kobayashi:2016xpl} for the case without torsion. In the case with torsion,  the certainty of instabilities {\it or} an eventual superluminality of the graviton was proven in  \cite{Mironov:2023wxn} for a restricted class of theories. We refer to the latter also as a "global stability" issue, because having "stability" would force an arguably unphysical and {\it potentially} unacceptable period of superluminality \cite{Adams:2006sv,Dubovsky:2005xd,Easson:2013bda,Creminelli:2022onn} (See, however, the discussion in \cite{Babichev:2007dw}). 

All in all, it has been argued that the global stability issues can be postponed and cured with other types of matter that could become relevant at other phases. However, there are generalizations of the No-Go theorems that hold for even more general modifications of GR \cite{Akama:2017jsa}, and furthermore, some analysis have suggested that violating the NEC potentially brings issues of some sort such as superluminality, even if singularities are successfully avoided \cite{Dubovsky:2005xd}. Thus, the global stability seems indeed a pressing issue that has to be solved concretely. Among the possible options, one is to consider altogether other theories such as Beyond Horndeski \cite{Mironov:2018oec,Mironov:2019mye}. In this case the equations of motion are of higher than second order, but there is no Ostrogradsky ghost by construction. Even for Horndeski theory there are very particular solutions to the global stability issue \cite{Cai:2016thi,Cai:2017tku,Cai:2017dyi,Creminelli:2016zwa,Kobayashi:2016xpl, Mironov:2022quk,Ageeva:2021yik}, but one is restricted to one of the following three options: either the model propagates no scalar perturbation about a nonsingular Friedmann-Lemaître-Robertson- Walker (FLRW) background - which may be unsatisfactory because we  do expect small deviations from FLRW on cosmological scales - or the scalar perturbation propagates about Minkowski spacetime \cite{Mironov:2022quk}, or one is forced to consider non conventional asymptotics, such as gravity being the strongest force in the past \cite{Kobayashi:2016xpl,Ageeva:2021yik}. Finally, substantially reconsidering Horndeski theory, now on a flat spacetime and with extra terms, fully {\it exchanging} curvature for torsion through the teleparallel connection, the usual No-Go theorems break \cite{Ahmedov:2023lot}. 

We suggest that another concrete solution - on the familiar  curved spacetime  - may be to simply lift some of the assumptions that are usually taken for Horndeski theory. Indeed, we relax the mathematically and physically unjustified\footnote{Indeed, there are motivations - beyond those considered in this paper - that suggest the relevance of torsion \cite{ Hehl:1976kj}. See also the discussions in \cite{Carroll:1994dq}. In fact, torsion had already been explored as a potential cure for the singularities of GR, but without addressing potential issues with global stability \cite{kopczynski1972non, Tafel:1973ayl, trautman1973structure, Hehl:1976kj}. } assumption of considering a spacetime {\it only} with curvature, as research in Horndeski theory historically developed. Because, as we show below, assuming vanishing torsion only helps to simplicity, but at the expense of "artificially" enabling the global stability issues. Indeed, it is remarkable that for such a general theory as Horndeski, a No-Go theorem for the linearization holds at all. It may be that in fact that the simplifying assumptions of the spacetime on which Horndeski theory was formulated are too restrictive given that so many criteria  - 1) nonsingular, 2) stable, 3) sub/ luminal cosmology- must be satisfied. We take 1) to 3) as a definition of  {\it "healthiness"} for what follows.

This work builds on the results for {\bf up to quartic} Horndeski theory with both curvature and torsion -namely, with up to $(\nabla^2 \phi)^2$ terms in the action- \cite{Mironov:2023wxn}, where a No-Go theorem for healthy cosmological evolution was formulated \footnote{We consider torsion in the second order, metric formalism. Namely, we assume from the start that the action can be written with a connection that can be expressed in terms of the Christoffel connection plus torsion \cite{Mironov:2023kzt,valenciavillegas}. See \cite{Ahmedov:2023lot,Ahmedov:2023num,helpin2020varying,helpin2020metric,dong2022constraining,davydov2018comparing,dong2022polarization,capozziello2023ghost,bahamonde2022symmetric,dialektopoulos2022classification,bernardo2021well,bahamonde2020post,bahamonde2020reviving,
bahamonde2019can} for other formalisms and modifications of Horndeski.}. In this paper, it is shown that once we include the more complex $\mathcal{L}_5$ part of the action -namely, with up to $(\nabla^2 \phi)^3$ terms-, a No-Go theorem cannot hold. We first argue that the {\it pressing issues} with the {\it simpler forms of} Horndeski theory -including the torsionless \cite{Libanov:2016kfc, Kobayashi:2016xpl} and the "up to quartic torsionful theory" in \cite{Mironov:2023wxn}- can be seen as simply {\it accidental to the assumptions}: namely, in the familiar Horndeski theory without torsion many of the coefficients of the quadratic action for the graviton are tightly related to the coefficients of the quadratic action for the scalar mode, and this is nothing more than the consequence of having taken {\it ad-hoc} simplifications. This link between coefficients ends up in the well-known contradictory requirements for the healthiness of the graviton and the scalar mode (No-Go theorems). A simple pragmatical solution is to lift the assumptions: Turning on torsion modifies the action for the graviton and the scalar mode in increasingly {\it divergent} ways, as more terms of the full Horndeski action are considered.  These modifications come from nontrivial tensor and scalar perturbations of torsion, which mix with the metric perturbations. The more mixing, the more the link between the graviton and scalar actions is broken. For instance, in  the "up to quartic torsionful theory" in \cite{Mironov:2023wxn} the mixing of more perturbations is enough to break the standard  No-Go theorem \cite{Libanov:2016kfc, Kobayashi:2016xpl}, thus ending up in a weaker form of this No-Go, where one can achieve stability and nonsingular cosmology, but with an arbitrarily short  superluminality of the graviton \cite{Mironov:2023wxn}. In this paper we show that if we lift further the "simplicity assumptions" and consider the complete form of Horndeski theory -including the  $\mathcal{L}_{5}$ part -  more torsion perturbations do not decouple, but rather they mix more with the metric perturbations. This ends up untangling the equations that must be imposed for the healthiness of both sectors. They become two sets of very different combinations of the Lagrangian functions $\mathcal{L}_{2}$ to $\mathcal{L}_{5}$. Thus, it is easier to independently satisfy these equations not only locally but also without meeting any contradictions at some time in the evolution.  Namely, reaching global stability. We show in this work an explicit toy model with  globally healthy cosmological solutions.

Furthermore,  {\it the gravitational waves  in all potentially healthy Horndeski gravities with torsion} follow a characteristic dispersion relation that is not common to any other simpler form of Horndeski theory about the spatially flat FLRW background. Their dispersion relation depends non-trivially on the wavelength. This follows because of the more torsion perturbations that do not decouple, which is necessary to avoid the No-Go's.

We proceed as follows: In section \ref{sec themodel} we define the model. In section \ref{sec graviton} we summarize the gravitational waves' characteristic features in all potentially healthy models.  

In section \ref{sec nonogo} we compare the situation regarding the global stability between three forms of Horndeski theory: {\bf (\ref{item disc1 no torsion})} without torsion, {\bf (\ref{item disc1 torsion upto4})} on a spacetime with torsion but without $\mathcal{L}_5$ (namely "up to quartic Horndeski-Cartan theory"), and {\bf (\ref{item disc1 torsion})} on a spacetime with torsion with the complete Horndeski action (which we denote as the full Horndeski-Cartan action). We explain how the No-Go holds for the  simpler forms of the theory {\bf (\ref{item disc1 no torsion})},{\bf (\ref{item disc1 torsion upto4})}, and how the usual analysis breaks for {\bf (\ref{item disc1 torsion})}, thus first  suggesting that with the full Horndeski-Cartan action {\bf (\ref{item disc1 torsion})} healthy solutions can be built. 

Finally, in section \ref{sec example} we show that a toy model exists, taken from within the general Horndeski-Cartan theories, which has a classical solution that is nonsingular,  stable and sub/luminal {\it at all times}, thus showing that a "No-Go theorem" cannot exist. We finalize with the conclusions in section \ref{sec conclusions}. 

In the Appendix \ref{secapp example} we report the construction details of the stable toy model, in \ref{secapp notation} we present details of the notation for torsion, and in \ref{secapp equations backgrounds}, \ref{secapp coefficients tensor modes}, \ref{secapp scalar sector} we show all necessary details of the quadratic actions for the tensor and scalar modes.

\section{The model: Horndeski with curvature and torsion}\label{sec themodel}

Horndeski theory is built on top of GR with four general functions $G_2,\, G_3,\, G_4,\, G_5$ that depend on a real scalar field $\phi$ and its first derivatives $X=-\frac{1}{2}\partial_\mu\phi\, \partial^\mu\phi $. These general functions appear in specific Lorentz invariant combinations of the second derivative term $(\tilde{\nabla}_\mu \tilde{\nabla}_\nu \phi)^p$  with $p\leq 3$, such that the equations of motion of all the fields are at most of second order, hence avoiding the Ostrogradsky ghost.

If we consider a spacetime with curvature and torsion, the basic block to build the second-order part of the Horndeski action $(\tilde{\nabla}_\mu \tilde{\nabla}_\nu \phi )$ is not symmetric under the exchange of indices $\mu \leftrightarrow \nu$. Thus, there are more ways to build Lorentz invariant combinations of the type $(\tilde{\nabla}_\mu \tilde{\nabla}_\nu \phi)^p$ and still keep the second order equations of motion. These multiple choices of Horndeski Lagrangians on a spacetime with torsion may be named by a handful of free parameters\footnote{Namely, for Horndeski with torsion with up to $\mathcal{L}_4$ there are two free parameters \cite{valenciavillegas}. Including $\mathcal{L}_5$ it is clear that many more free parameters are possible, but it has not been classified.}, which for different choices may lead to fundamentally different dynamics within the family of Horndeski-Cartan theories \cite{Mironov:2023kzt,valenciavillegas}. 

Currently, it is known that the most physically compelling type\footnote{Namely, the theory that besides a graviton also propagates a scalar mode with a relativistic dispersion relation. This corresponds to a particular choice of parameters of a two-parameter family of (up to quartic $G_4$) Horndeski-Cartan theories \cite{Mironov:2023kzt,valenciavillegas}.} of Horndeski-Cartan theories with only $G_2,\, G_3,\,G_4$ functions ("up to quartic") eventually suffers of a pathology in the classical solutions at some time in the evolution of the universe, or the graviton needs to become superluminal for at least an arbitrarily short time \cite{Mironov:2023wxn}. More precisely, in up to quartic theories that can be written as the $\mathcal{L}_2\,+\, \mathcal{L}_3\,+\, \mathcal{L}_4 $ part of the action (\ref{eqn action}), namely, with up to $(\tilde{\nabla}_\mu \tilde{\nabla}_\nu \phi)^2$ Lorentz invariant combinations, it is not possible to obtain a  FLRW cosmology that is always subluminal, nonsingular and stable. Denoting torsionful\footnote{In particular, note that the Einstein tensor $\tilde{G}^{\mu\nu}$ is {\it not} symmetric due to torsion.} quantities with a "tilde" $(\, \tilde{}\,)$, and $\partial G_A/\partial X=:G_{A,X}$,
\begin{eqnarray} 
\mathcal{S}&=&\int \text{d}^4 x\, \left(\mathcal{L}_2\,+\, \mathcal{L}_3\,+\, \mathcal{L}_4\,+\, \mathcal{L}_5 \right)\,,\label{eqn action}\\
\mathcal{L}_2&=& G_2\,,\\
\mathcal{L}_3&=& -G_{3}\, {\tilde{\nabla}}_\mu{\tilde{\nabla}}^\mu\phi\,,\\
\mathcal{L}_4&=& \text{\resizebox{0.38\textwidth}{!}{$G_4\,\tilde{R}+G_{4,X}\left(\left(\tilde{\nabla}_\mu\tilde{\nabla}^\mu\phi\right)^2-\left(\tilde{\nabla}_\mu\tilde{\nabla}_\nu\phi\right) \tilde{\nabla}^\nu\tilde{\nabla}^\mu\phi \right)$}}.
\end{eqnarray}

In this letter we show that {\it the situation changes drastically} if we also consider the remaining structure of Horndeski theory, namely with the $G_5$ part which contains up to $(\tilde{\nabla}_\mu \tilde{\nabla}_\nu \phi)^3$ Lorentz invariant combinations. 

Specifically, let us take the complete action (\ref{eqn action}) and, among the many possible terms that a spacetime with curvature and torsion allows, we consider only the following Lorentz contractions of $(\tilde{\nabla}_\mu \tilde{\nabla}_\nu \phi)^3$ terms
\begin{eqnarray}
&&\mathcal{L}_{5}=\,G_5 \,\tilde{G}^{\mu\nu}\, \tilde{\nabla}_\mu \tilde{\nabla}_\nu\phi-\frac{1}{6}\,G_{5,\,X}\,\Big((\tilde{\nabla}_\mu \tilde{\nabla}^\mu \phi)^3 \label{eqn L5}\\
&+& (\tilde{\nabla}_\nu\tilde{\nabla}_\rho\phi)\Big(2 (\tilde{\nabla}^\mu \tilde{\nabla}^\nu\phi) \tilde{\nabla}^\rho \tilde{\nabla}_\mu\phi-3(\tilde{\nabla}_\mu \tilde{\nabla}^\mu\phi) \tilde{\nabla}^\rho \tilde{\nabla}^\nu\phi\Big) \Big)\,. \nonumber
\end{eqnarray}
Any other choice of contractions of Lorentz indices in both $\mathcal{L}_4$ and $\mathcal{L}_5$ leads to fundamentally different dynamics\footnote{See for instance \cite{Mironov:2023kzt} for the discussion in $\mathcal{L}_4$. In general they may have less constraints, making them algebraically more difficult to analyze. Their analysis is left for future work.}, even if the equations of motion are kept of second order. 

The specific {\it subclass} (\ref{eqn action}) - (\ref{eqn L5}) which we have chosen within the family of all Horndeski-Cartan theories will be enough to show that there exist at least some Horndeski-Cartan theory that can support an all time stable, nonsingular and subluminal cosmology.

\section{Stable cosmologies with a distinctive graviton}\label{sec graviton and scalar}

We examine the stability of the perturbations against a spatially flat FLRW background. As in GR there is no dynamical vector perturbation. And similar as in Horndeski without torsion, there are the usual two polarizations of the graviton and a scalar mode. However, their speed and stability are markedly different for the full theory with torsion:

\subsection{Modified graviton}\label{sec graviton}

For all actions of the type (\ref{eqn action}), containing the $G_5$ structure on a torsionful spacetime, the dispersion relation of gravitational waves ($\omega$) on the FLRW background is such that the speed generally depends on the wavelength\footnote{This is not that surprising given that the cosmological background spontaneously breaks Lorentz invariance. Indeed, this is common in some derivatively coupled scalars, such as in ghost condensates, which usually have a non relativistic scalar dispersion relation \cite{arkani2004ghost}.},
\begin{equation}
\omega^2 =\frac{f_0+\vec{p}\, {}^2 \,f_1}{f_2+\,\vec{p}\, {}^2\,f_3}\,\vec{p}\, {}^2\,, \label{eqn cgw}
\end{equation}
where we have written all the dependance on momentum ($\vec{p}$) explicitly, and $f_0,\,f_1,\,f_2,\, f_3$ depend on the Lagrangian functions. They are given in the Appendix \ref{secapp coefficients tensor modes}.

For our analysis it is appropriate to take the short wavelength approximation and consider the speed of gravitational waves as $\vec{p}\, {}^2\rightarrow \infty$,
\begin{eqnarray}
c_g^2&=&\frac{f_1}{f_3}\,.\label{eqn cg2}
\end{eqnarray}

This peculiar dispersion relation for the graviton  prevails for all Horndeski theories with torsion that contain $G_5$, irrespectful of the order of contraction of Lorentz indices in any term of the action (\ref{eqn action}). Indeed, it is useful to notice in Eq. (\ref{eqn cgw}), $f_1\propto f_3 \propto G_5^2$. Thus, we can already  conclude that {\it all potentially healthy} models of Horndeski on a space-time with Torsion will be characterized by the dispersion relation (\ref{eqn cgw}), because they must contain the $G_5$ part. Indeed, in \cite{Mironov:2023wxn} it was proven that  a No-Go theorem holds for the simpler form of Horndeski with torsion without $G_5$ (See the analysis below).

Let us now see how this new graviton with the distinctive Eq. (\ref{eqn cgw}) helps to the stability of Horndeski models in comparison to simpler forms of the theory:

\subsection{Dodging the No-Go}\label{sec nonogo}

Let us briefly see how the dispersion relation in Eq. (\ref{eqn cgw}) comes to be markedly different in comparison to simpler forms of Horndeski theory. It is obtained as follows: Defining torsion as the difference between connections $T^{\rho}{}_{\mu\nu}=\tilde{\Gamma}^\rho_{\mu\nu}-\tilde{\Gamma}^\rho_{\nu\mu}$, we note that by its antisymmetry in lower indices it has $24$ independent components, of which we count two two-component tensor perturbations $T^{\scalebox{0.5}{(1)}}_{ij},\, T^{\scalebox{0.5}{(2)}}_{ij} $ about the spatially flat FLRW background\footnote{See Appendix \ref{secapp notation} for more details on our notation for torsion and Appendix \ref{secapp decomposition and lin} for a detailed decomposition of perturbations and background fields. We follow the same notation as in  \cite{Mironov:2023kzt}.}. Hence, denoting with $h_{ij}$ the tensor perturbation of the metric, the action (\ref{eqn action}) for all tensor perturbations about the FLRW background (in conformal time), in momentum space takes the form 
\begin{eqnarray}
&&\mathcal{S}_{\tau}= \int\, \textrm{d}\eta\,\textrm{d}^3p \,\,\Big( b_{1}\,(\dot{h}_{ij})^2+b_{2} \, \vec{p}\, {}^2 (h_{ij})^2+ b_3 (h_{ij})^2 \label{eqn Ltensor1}  \\
&& + \Big( c_{1}\, \vec{p}\, {}^2( T^{\scalebox{0.5}{(2)}}_{ij})^2 +c_2 \,h_{ij}\, T^{\scalebox{0.5}{(1)}}_{ij} + c_3  \,\dot{h}_{ij}\, T^{\scalebox{0.5}{(1)}}_{ij} + c_{4}\, (T^{\scalebox{0.5}{(1)}}_{ij})^2 \Big) \nonumber\\
&& + \vec{p}\, {}^2\Big(d_1  \, T^{\scalebox{0.5}{(1)}}_{ij} +d_2  \,\dot{h}_{ij}\,  +d_3  \,h_{ij}\, \Big)\,  T^{\scalebox{0.5}{(2)}}_{ij}+ d_4 \, \vec{p}\, {}^2 h_{ij}\, T^{\scalebox{0.5}{(1)}}_{ij}\Big)\,,\nonumber
\end{eqnarray}
where $b_A,\,c_A,\,d_A$ depend on three background fields: the scale factor of the FLRW metric $a(\eta)$, the Horndeski scalar, which in the context of linearized expressions we also denote as $\phi(\eta)$ and a non-trivial torsion background $x(\eta)$, which, however, can be solved in terms of $a(\eta),\, \phi(\eta)$ (See Appendices \ref{secapp decomposition and lin} and \ref{secapp equations backgrounds}).

The situation for the gravitational waves compares as follows between different forms of Horndeski theory: 

{\bf {\bf (\ref{item disc1 no torsion})} without torsion,

{\bf (\ref{item disc1 torsion upto4})} on a spacetime with torsion with the action (\ref{eqn action}) but without $\mathcal{L}_5$ (which we refer as "up to quartic Horndeski-Cartan theory"), and 

{\bf (\ref{item disc1 torsion}) on a spacetime with torsion with the action (\ref{eqn action}) including $\mathcal{L}_5$} (which we denote as the full Horndeski-Cartan action}\footnote{Let us however note that one could still consider additional contractions of Lorentz indices.}).

\begin{enumerate}[{\bf (i)}]
\item{On a spacetime {\it without} torsion only the terms $b_A$ in the first line in (\ref{eqn Ltensor1}) contribute. In this case one recognizes that the graviton is massless $(b_3=0)$ after using the equations for the background fields.} \label{item disc1 no torsion}
\item{On a spacetime with torsion but without $\mathcal{L}_5$, the first and second lines in (\ref{eqn Ltensor1}) do contribute, but the third does not (namely, $d_A=0$). Again the graviton turns out to be massless, but its speed is modified due to its coupling $c_2,\, c_3$ to one of the torsion perturbations, $T^{\scalebox{0.5}{(1)}}_{ij} $. The essential aspect is that in this case  $T^{\scalebox{0.5}{(2)}}_{ij} $ fully decouples, because $d_1=d_2=d_3=0$ (namely, its equation gives $c_1\, T^{\scalebox{0.5}{(2)}}_{ij}\equiv 0$), and as a consequence the modification to the graviton of the up to quartic Horndeski-Cartan theory is not as marked as in (\ref{eqn cgw}), as reported in \cite{Mironov:2023kzt,Mironov:2023wxn,valenciavillegas}.} \label{item disc1 torsion upto4}
\item{On the other hand, for the full Horndeski-Cartan action (\ref{eqn action}), the third line in (\ref{eqn Ltensor1}) also contributes and now $T^{\scalebox{0.5}{(2)}}_{ij} $ no longer decouples. This is the critical difference. Its equation of motion solves torsion as
\begin{eqnarray}
T^{\scalebox{0.5}{(2)}}_{ij}=-\frac{1}{2\,c_1} \Big(d_1  \, T^{\scalebox{0.5}{(1)}}_{ij} +d_2  \,\dot{h}_{ij}\,  +d_3  \,h_{ij}\Big)\,.\label{eqn T2sol}
\end{eqnarray}
With this solution back in (\ref{eqn Ltensor1}) we now obtain a $-\frac{d_1^2}{4\,c_1} \vec{p}\, {}^2 $ momentum contribution to $(T^{\scalebox{0.5}{(1)}}_{ij})^2 $, besides the $c_4$ term. Indeed, now there is a term of the form $-\frac{1}{4 \,c_1}(-4\,c_1\,c_{4}+ \vec{p}\, {}^2\, d_1^2)\, (T^{\scalebox{0.5}{(1)}}_{ij})^2 $. This lies at the core of the peculiar dispersion relation of the graviton, because $T^{\scalebox{0.5}{(1)}}_{ij} $ also couples to $\dot{h}_{ij}$ and $h_{ij}$, and, if $d_1\propto G_5$ is not identically zero, then there will be a  $\frac{1}{\mathcal{O}(\vec{p}\, {}^2)}$ contribution to the kinetic and gradient terms for the graviton.

Indeed, skipping unimportant details for this discussion, the torsion perturbation $T^{\scalebox{0.5}{(1)}}_{ij} $ can be finally solved in (\ref{eqn action}) in terms of $h_{ij},\, \dot{h}_{ij}$, and critically, with terms of order $\frac{1}{\mathcal{O}(\vec{p}\, {}^2)}$, provided $c_1,\,c_2,\,c_3$ are not identically zero,
\begin{eqnarray}
T^{\scalebox{0.5}{(1)}}_{ij}=\frac{1}{f_2+ \vec{p}\, {}^2 \,f_3} \left( \left(2\,c_1\,c_3-\,\vec{p}\, {}^2\,d_1\,d_2\right)\dot{h}_{ij}+ \left(2\,c_1\,c_2-\,\vec{p}\, {}^2\,\left(d_1\,d_3-2\,c_1\,d_4\right)\right){h}_{ij}\right) \,,\label{eqn T1sol}
\end{eqnarray}
with $f_2= -4\,c_1\,c_4 \,,\, f_3=d_1^2$. Finally, using the solutions Eq. (\ref{eqn T1sol}), (\ref{eqn T2sol}) we can write the action (\ref{eqn Ltensor1}) for the usual graviton with two polarizations, as:
\begin{eqnarray}
\mathcal{S}_{\tau}= \int\, \textrm{d}\eta\,\textrm{d}^3p\,a^4\, \left[\frac{1}{2\,a^2}\,\frac{1}{f_2+ \vec{p}\, {}^2 \,f_3} \,\left(\bar{\mathcal{G}}_\tau\,(\dot{h}_{ij})^2
-\, \vec{p}\, {}^2\,\bar{F}_\tau\,(h_{ij})^2\right)\right]\,
\label{eqn Ltensor2}
\end{eqnarray}
with,
\begin{equation}
\begin{array}{cc}
\bar{\mathcal{G}}_\tau =\frac{2}{a^2} c_1\,(c_3^2-4\,b_1\,c_4)\,, & \bar{\mathcal{F}}_\tau(\vec{p}\, {}^2) =\frac{\bar{f}_0+\vec{p}\, {}^2 \, \bar{f}_1}{f_2+ \vec{p}\, {}^2 \,f_3}
\end{array}
\end{equation}
from which we identify the dispersion relation (\ref{eqn cgw}) with $f_0=\bar{f}_0/\bar{\mathcal{G}}_\tau,\, f_1=\bar{f}_1/\bar{\mathcal{G}}_\tau $. It is convenient to also define\footnote{The global factor in the part of the action for the tensor modes, of order $\frac{1}{\mathcal{O}(\vec{p}\, {}^2)}$, may indicate nothing more than that at very  high momenta the theory is strongly coupled, which we already expect from this effective theory. However, it is clear that the speed of gravitational waves is well defined in any case. Now, whether this presents a danger in the perturbative expansion falls beyond the reach of the first approximation in this paper, restricted to linear order.}
\begin{equation}
\begin{array}{cc}
\mathcal{G}_{\tau} =\frac{\bar{\mathcal{G}}_{\tau}}{f_2+ \vec{p}\, {}^2 \,f_3}\,, & \mathcal{F}_{\tau} =\frac{\bar{\mathcal{F}}_{\tau}}{f_2+ \vec{p}\, {}^2 \,f_3}\,.
\end{array}\label{eqn GtFt}
\end{equation}}\label{item disc1 torsion}
\end{enumerate}

Now, in most forms of Horndeski theory the stability or no-ghost requirements for both the tensor and scalar sectors are contradictory. The issue is that  the coefficients in the action for the tensor and scalar sectors are tightly related. In up to quartic Horndeski-Cartan also the subluminality of the graviton is part of the inconsistent assumptions. Let us see how the tight relation between tensor and scalar actions is broken for the full Horndeski action with torsion (\ref{eqn action}).

The scalar sector for the action (\ref{eqn action}) can be brought to a form that is typical of Horndeski theories in the unitary gauge, after integrating out all scalar perturbations of torsion, because they are non dynamical (We show this in detail in the Appendix \ref{secapp scalar sector}). Namely:
\begin{dmath}
\mathcal{S}_{s}\,\text{$=$}\, \int\, \textrm{d}\eta\,\textrm{d}^3x\, a^4 \left(- 3\,\frac{ \, \bar{\mathcal{G}}_{\mathcal{S}}}{a^2}\,\dot{\psi}^2 \,+ \frac{ \,\bar{\mathcal{F}}_{\mathcal{S}}}{a^2}\, \,({{\partial_i \psi}})^2 +6\,\frac{ \Theta}{a}\,\alpha\,\dot{\psi}\,+ 2\, \frac{T}{a^2} \,{{\partial_i \alpha}} \,{{\partial_i \psi}} \,+ 2\frac{{{\partial_i \partial_i B}}}{a^2}\,\left(\,a\, \Theta \,{{\alpha}}\,-\, \, \bar{\mathcal{G}}_{\mathcal{S}}\, \,{\dot{\psi}}\right) \,+\,\Sigma\,\alpha^2 \right)\,\,,\nonumber\\
\label{eqn Lscalar}
\end{dmath}
where $\psi,\, \alpha$ and $B$ are scalar perturbations for the metric, and the coefficients $\bar{\mathcal{G}}_{\mathcal{S}},\,\bar{\mathcal{F}}_{\mathcal{S}},\, \Theta,\, T,\, \Sigma$, given in the Appendix \ref{secapp scalar sector}, depend on the backgrounds $a(\eta),\, \phi(\eta)$, but they {\it do not depend on spatial momentum}. Or, using the constraint equation $\alpha=\frac{\bar{\mathcal{G}}_{\mathcal{S}}}{a\,\Theta}\,\dot{\psi}$ imposed by the Lagrange multiplier $B$ in Eq. (\ref{eqn Lscalar}), the action for a single dynamical scalar  finally reads
\begin{eqnarray}
\mathcal{S}_{s}= \int\, \textrm{d}\eta\,\textrm{d}^3x \,a^4\,\left(\frac{1}{a^2}\, \mathcal{G}_{\mathcal{S}}\,\dot{\psi}^2-\, \frac{1}{a^2}\, \mathcal{F}_{\mathcal{S}}\,(\partial_i \psi)^2\right)\,,\label{eqn ql finalStep}
\end{eqnarray}
where,
\begin{eqnarray}
\begin{array}{cc}
\mathcal{G}_{\mathcal{S}}= 3\, \bar{\mathcal{G}}_{\mathcal{S}} +\frac{\bar{\mathcal{G}}_{\mathcal{S}} ^2\,\Sigma}{\Theta^2} \label{eqn Gs}\,, \,&
\mathcal{F}_{\mathcal{S}}= \frac{1}{a^2}\frac{\text{d}N}{\text{d}\eta} -\bar{\mathcal{F}}_{\mathcal{S}}\,, \label{eqn Fs}
\end{array}
\end{eqnarray}
with $c_{\mathcal{S}}^2=\mathcal{F}_{\mathcal{S}}/\mathcal{G}_{\mathcal{S}}$ the speed squared of the scalar mode, and where we have defined
\begin{equation}
N=: \frac{a\, \bar{\mathcal{G}}_{\mathcal{S}} \, T}{\Theta} \,,\label{eqn N}
\end{equation}
which is a quantity of major importance for the discussions below.

Now, let us see how the argument on the stability compares between the different forms of Horndeski theory {\bf (\ref{item disc1 no  torsion})} to {\bf (\ref{item disc1 torsion})} defined before: assuming "normal" asymptotics -more precisely, assuming no strong gravity in the asymptotic past or future $\mathcal{F}_{\tau}(\eta)>b_2>0$ as $\eta\rightarrow \pm\infty$ - the following compelling assumptions (A), (B) are already mutually inconsistent in the case {\bf (\ref{item disc1 no  torsion})} of Horndeski theory  without torsion\footnote{In this discussion we remain within the framework of general Horndeski theories not defined by the particular eqution $\Theta\equiv 0$, which solves the issues in the torsionless Horndeski theory at the expense of loss of generality, having only two options:  either the model propagates no scalar perturbation about a nonsingular FLRW background, or the scalar perturbation propagates about Minkowski spacetime  \cite{Mironov:2022quk}.}:
\begin{enumerate}[A)]
\item{nonsingular cosmological solution (a lower bound on the scale factor $a(\eta)>b_1>0$),}\label{eqn nonsingular cosmology}
\item{the graviton and the scalar mode are not ghosts and they suffer no gradient instabilities, 
$\mathcal{G}_{\tau}>0 \,, \mathcal{F}_{\tau}>0 \,,  \mathcal{F}_{\mathcal{S}}>0 \,,  \mathcal{G}_{\mathcal{S}}>0\,. $}\label{eqn stability conditions}
\end{enumerate}
In the case {\bf (\ref{item disc1 torsion upto4})} of torsionful, up to quartic Horndeski theory one finds an inconsistency if together with the assumptions (\ref{eqn nonsingular cosmology}, \ref{eqn stability conditions}) we also assume that
\begin{enumerate}[C)]
\item{the graviton is always sub/ luminal ($c_g^2\leq 1$).}\label{eqn subluminality condition} 
\end{enumerate}

\begin{figure}[t]
 \centering
\begin{subfigure}{0.5\textwidth}
\centering
  \includegraphics[width=0.8\linewidth]{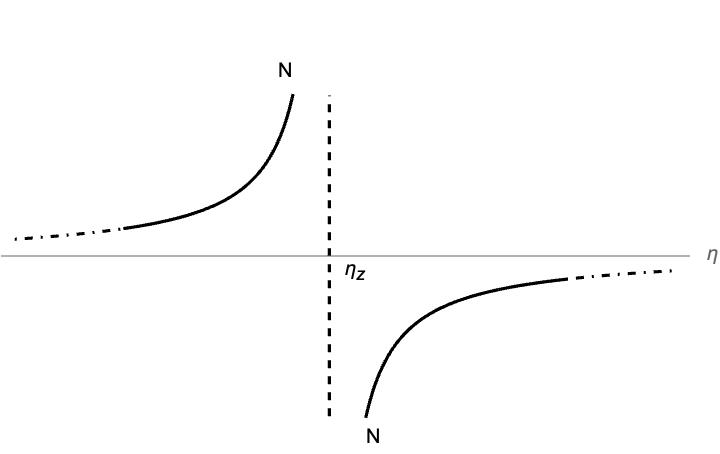}
 \caption{$N$ monotonous growing around any zero $\eta_z$.} \label{fig aroundOneZero}
\end{subfigure}%
\begin{subfigure}{0.5\textwidth}
\centering
  \includegraphics[width=1\linewidth]{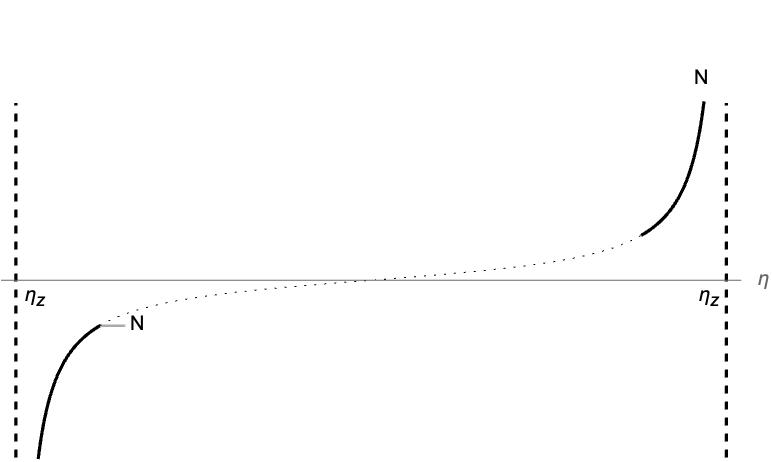}
 \caption{$N$ in between any two zeros $\eta_z$. $N$ is continuous in that interval and hence vanishes at some point.} \label{fig inBetweenTwoZeros}
\end{subfigure}\caption{Behavior of $N(\eta)$ around zeros of $\Theta$ (denoted as $\eta_z$).}
\end{figure}

\begin{figure}[t]
 \centering
  \includegraphics[width=0.5\linewidth]{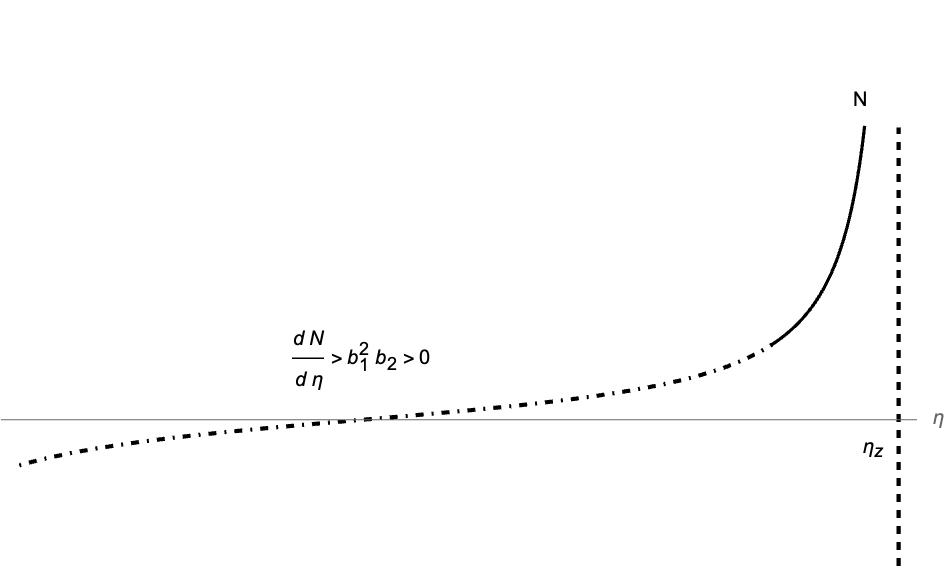}
 \caption{$N$ in a semi-infinite interval $(-\infty,\eta_z)$. Similarly $N$ vanishes at some time in an interval $(\eta_z,\infty)$ or if there are no zeros $\eta_z$ at any time of the evolution, because $N$ cannot have horizontal asymptotes $\frac{\text{d}N}{\text{d}\eta} >b_1^2\, b_2>0$ with $b_1$ and $b_2$ positive constants, as $\eta\rightarrow \pm \infty$.}\label{fig inSemiInfiniteInterval}
\end{figure}

Indeed, following the argument initially shown in  \cite{Rubakov:2016zah,Libanov:2016kfc, Kobayashi:2016xpl} and then extended to the case with torsion in \cite{Mironov:2023wxn}, we can see the inconsistency in both cases {\bf (\ref{item disc1 no torsion})}, {\bf (\ref{item disc1 torsion upto4})}, as follows:

\begin{enumerate}[\indent{}]
\item{On one hand $N$ must not vanish provided the assumption of a nonsingular cosmology together with the no-ghost condition and sub/luminality of the graviton: namely, in {\bf (\ref{item disc1 no torsion})} $\bar{\mathcal{G}}_{\mathcal{S}} =T= \mathcal{G}_{\tau} $ and $\bar{\mathcal{F}}_{\mathcal{S}}= \mathcal{F}_{\tau}$, hence $N \propto a\, \mathcal{G}_{\tau}^2\neq 0$. And in {\bf (\ref{item disc1 torsion upto4})},  $\bar{\mathcal{G}}_{\mathcal{S}} = \mathcal{G}_{\tau}>0 $, $\bar{\mathcal{F}}_{\mathcal{S}}= \mathcal{F}_{\tau}$ and $T= \mathcal{F}_{\tau} \,(c_g^2-2)<0$ hence $N \propto a\, \mathcal{G}_{\tau}\, T\neq0$.

On the other hand, the stability requirement for the scalar mode $\mathcal{F}_{\mathcal{S}}>0$ in (\ref{eqn stability conditions}) tells, firstable, that the function $N$ is monotonous increasing $\frac{\text{d}N}{\text{d}\eta} >a^2\, \mathcal{F}_{\tau}>0$. This reveals the behavior of $N$ around any isolated zeros of $\Theta$, which we denote as $\eta_z$, as in Figure \ref{fig aroundOneZero}. In particular it implies that in between any two zeros of $\Theta$, $N$ must vanish as in Figure \ref{fig inBetweenTwoZeros}. Secondly, $\mathcal{F}_{\mathcal{S}}>0$ also tells that the slope of $N$ is bounded from below asymptotically: $\frac{\text{d}N}{\text{d}\eta} >b_1^2\, b_2>0$ as $\eta\rightarrow \pm \infty$. This implies that $N$ will also vanish in any infinite interval - even if $\Theta$ never vanishes - because $N$ cannot have a horizontal asymptote. See for instance Figure \ref{fig inSemiInfiniteInterval} in the case of the left-most zero of $\Theta$. This contradicts the last paragraph, hence, the physically compelling assumptions in each case {\bf (\ref{item disc1 no torsion})}, {\bf (\ref{item disc1 torsion upto4})} are mutually inconsistent. A detailed proof is given in \cite{Libanov:2016kfc,Kobayashi:2016xpl,Mironov:2019fop,Mironov:2023wxn}.\bigskip}
\end{enumerate}

Finally, in the case {\bf (\ref{item disc1 torsion})} the situation is markedly different. Considering the expression (\ref{eqn N}) for $N$, let us see that it can vanish in many ways: first, notice that in general the condition in (B) $\mathcal{G}_{\tau}>0$ does not constrain the sign of $\bar{\mathcal{G}}_{\mathcal{S}}$, because $\bar{\mathcal{G}}_{\mathcal{S}} \neq \mathcal{G}_{\tau}(\vec{p}\,{^2}) $ \footnote{Although the notation for $\bar{\mathcal{G}}_{\mathcal{S}}$ and $\mathcal{G}_{\mathcal{S}}$ is similar, they are very different functions (See Eqn. (\ref{eqn Gs}), and note that $\Sigma$ is a complicated function of the Lagrangian functions.). So, the condition in (B) $\mathcal{G}_{\mathcal{S}}>0$ does not constrain the sign of $\bar{\mathcal{G}}_{\mathcal{S}}$.}. This is obvious because while the latter depends on momentum, the former does not. Moreover, close inspection of the expression for $T$ in Eq. (\ref{eqn TG5}) in the Appendix \ref{secapp scalar sector}, shows that the latter is not related in a simple way to the speed of the graviton -defined only in the short wavelength {\it  approximation} (\ref{eqn cg2})-, which is the key aspect that would usually relate the stability to superluminality in up to quartic Horndeski-Cartan theory {\bf (\ref{item disc1 torsion upto4})}. This is not surprising, because the dispersion relation in Eq. (\ref{eqn cgw}) depends in a complicated way on momentum, while the coefficient $T$ in the scalar sector is independent of momentum.

Therefore, the usual assumptions to have a classically healthy solution (A)-(C) at some momenta -in particular $\mathcal{G}_{\tau}(\vec{p}\,{^2})>0$ and $c_g^2\leq 1$ - do not restrict $\bar{\mathcal{G}}_{\mathcal{S}}$ nor $T$, and one could potentially design a theory where (A)-(C) hold, and $\bar{\mathcal{G}}_{\mathcal{S}} $ or $T$ vanish at some time, and hence $N$ vanishes too. All in all, {\it if there exists a No-Go argument it must be a momentum dependent statement. A general statement fails because  the action for the tensor and scalar modes are modified in substantially diverging ways by torsion.} 

Now, it is important to mention a potentially special case. For that, let us  first notice the general identity 
\begin{equation}
\bar{\mathcal{G}}_{\tau}=f_2\, \bar{\mathcal{G}}_{\mathcal{S}}\,,\label{eqn idGtGs}
\end{equation}
which is still not a strong link between the tensor and scalar modes at all momenta - even if $\mathcal{G}_{\tau}>0$ and $\bar{\mathcal{G}}_{\tau}>0$ by (B) - because the sign of  $f_2(\eta)$ is in principle not constrained. However, by the relation (\ref{eqn idGtGs}) and the form of $\mathcal{G}_{\tau} $ in Eqn. (\ref{eqn GtFt}) it is clear that at low momentum a closer relation between the tensor and scalar modes arises\footnote{This is not that surprising given that in Eq. (\ref{eqn cgw}), the low momentum limit $\vec{p} \, {} ^2 \rightarrow 0$ is equivalent to the limit $G_5\rightarrow 0$.  Namely, the low momentum limit recovers a dispersion relation of the graviton similar to the form of "up to quartic" theories (let us recall that $f_1\propto G_5^2$ and $f_3\propto G_5^2$).}: namely as $\vec{p}\, {}^2\rightarrow 0$ we find
\begin{equation}
\begin{array}{cc}
\mathcal{G}_{\tau}(\vec{p} \, {} ^2=0)= \bar{\mathcal{G}}_{\mathcal{S}}\,,&
\mathcal{F}_{\tau}(\vec{p} \, {} ^2=0)= \bar{\mathcal{F}}_{\mathcal{S}}\,.
\end{array}\label{eqn linkTtoSlowP}
\end{equation}

Thus,  according to the inequalities (\ref{eqn stability conditions}) the no-ghost and stability of the tensor modes in the low momentum constrain the sign of $\bar{\mathcal{G}}_{\mathcal{S}} $ and $\bar{\mathcal{F}}_{\mathcal{S}} $ of the scalar mode. Evidently this reduces the ways in which $N$ can vanish at low momentum, but let us recall that $T$ is momentum independent and as we show in the Appendix \ref{secapp scalar sector} it may vanish in many different ways, thus making unplausible an exact No-Go even in this low momentum case. Indeed, we show an explicit counterexample to a No-Go in the next section. 

\begin{figure}[t]
 \centering
\begin{subfigure}{0.5\textwidth}
\centering
  \includegraphics[width=1\linewidth]{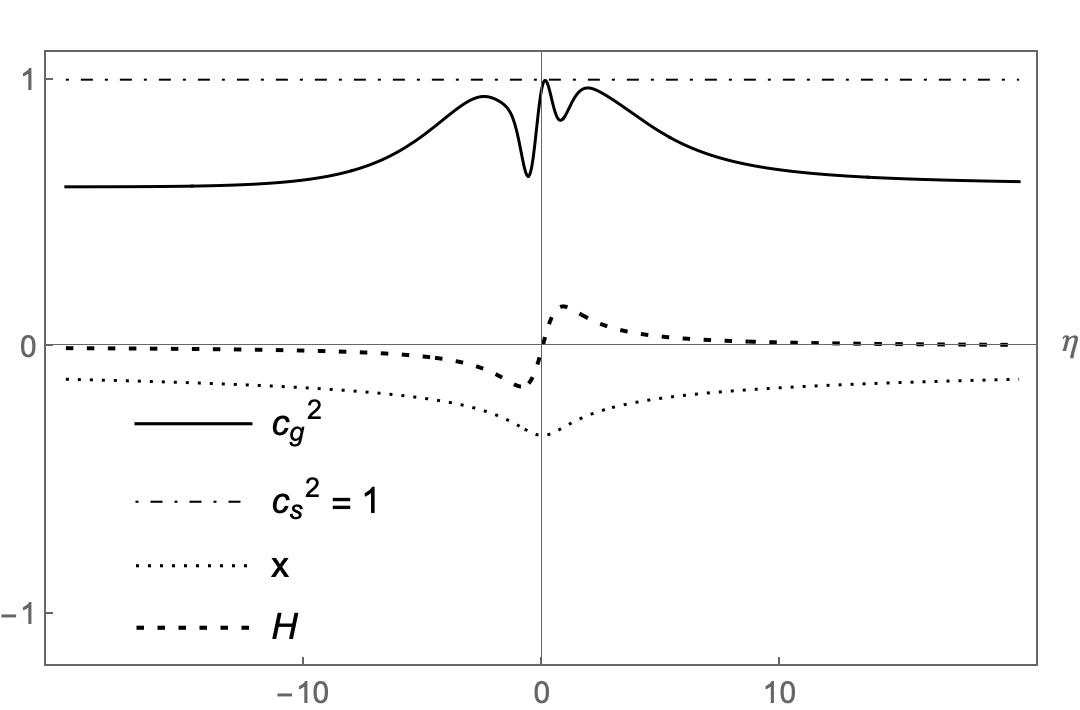}
 \caption{Speed of  gravitational waves and of the scalar mode. The Hubble showing a short bounce phase centered at $\eta=0$ and of width $\tau=1$. The isotropic torsion background, $x(\eta)$.} \label{}
\end{subfigure}%
\begin{subfigure}{0.5\textwidth}
\centering
  \includegraphics[width=1\linewidth]{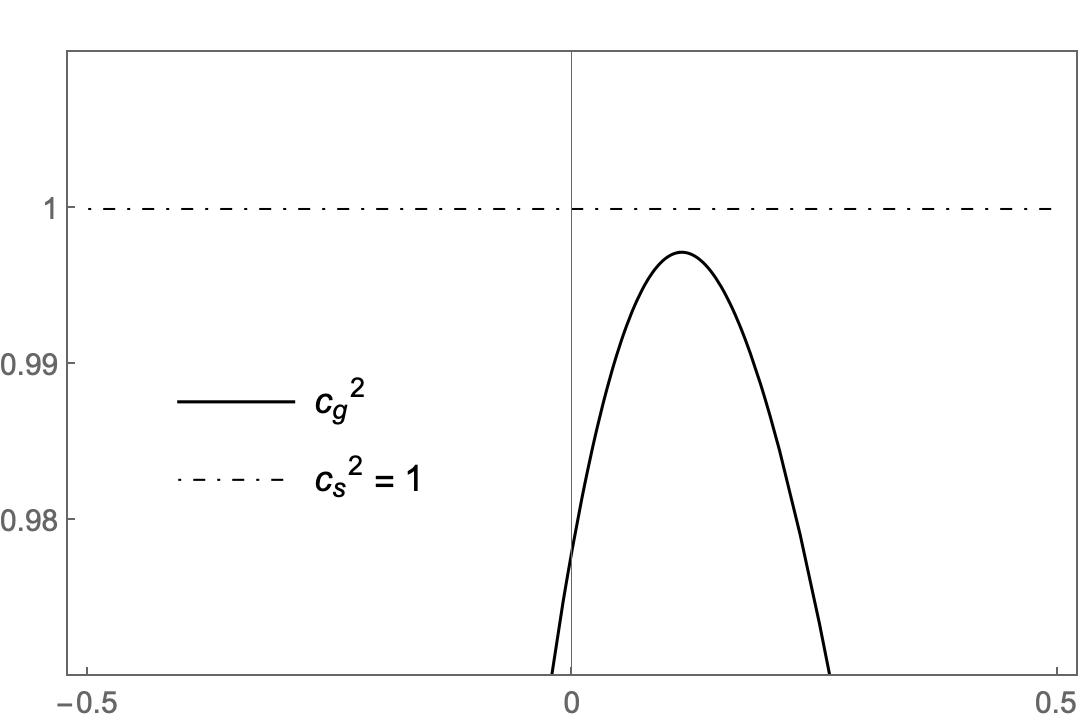}
 \caption{Detail of the speed of gravitational waves close to the bounce phase.} \label{}
\end{subfigure}\caption{Physically relevant quantities.}\label{fig speeds}
\end{figure}

\subsection{A counter example to a "No-Go": an all time stable, nonsingular and subluminal toy model}\label{sec example}

We support the previous discussion showing that a model exists, taken from within the general Horndeski-Cartan theories (with $G_5$) (\ref{eqn action}), which has a classical solution that is nonsingular,  stable and sub/luminal {\it at all times}, thus showing that {\bf a "No-Go theorem" cannot exist}.

Our sole intention is to show that there are such potentially interesting models. Thus, our criteria is limited to the latter and whether further features can be achieved - such as physically interesting asymptotics - is left as an open question.

The strategy to build the model is to take an Ansatz for the scalar potentials $G_2$ to $G_5$, as an expansion in powers of $X$, with coefficients that are general functions of $\phi$. Then, assuming a set of physically compelling solutions for a bouncing cosmology, $H$ and $\phi$ (Eqns. (\ref{eqn sol backgroundsM}), (\ref{eqn minimal highp})), we work backwards to solve the free functions in the Ansatz such that the equations of motion of the background fields are satisfied. In this section we summarize the key features of the model. The details are relegated to the Appendix \ref{secapp example}. 

\begin{figure}[b]
 \centering
\begin{subfigure}{0.5\textwidth}
\centering
  \includegraphics[width=1\linewidth]{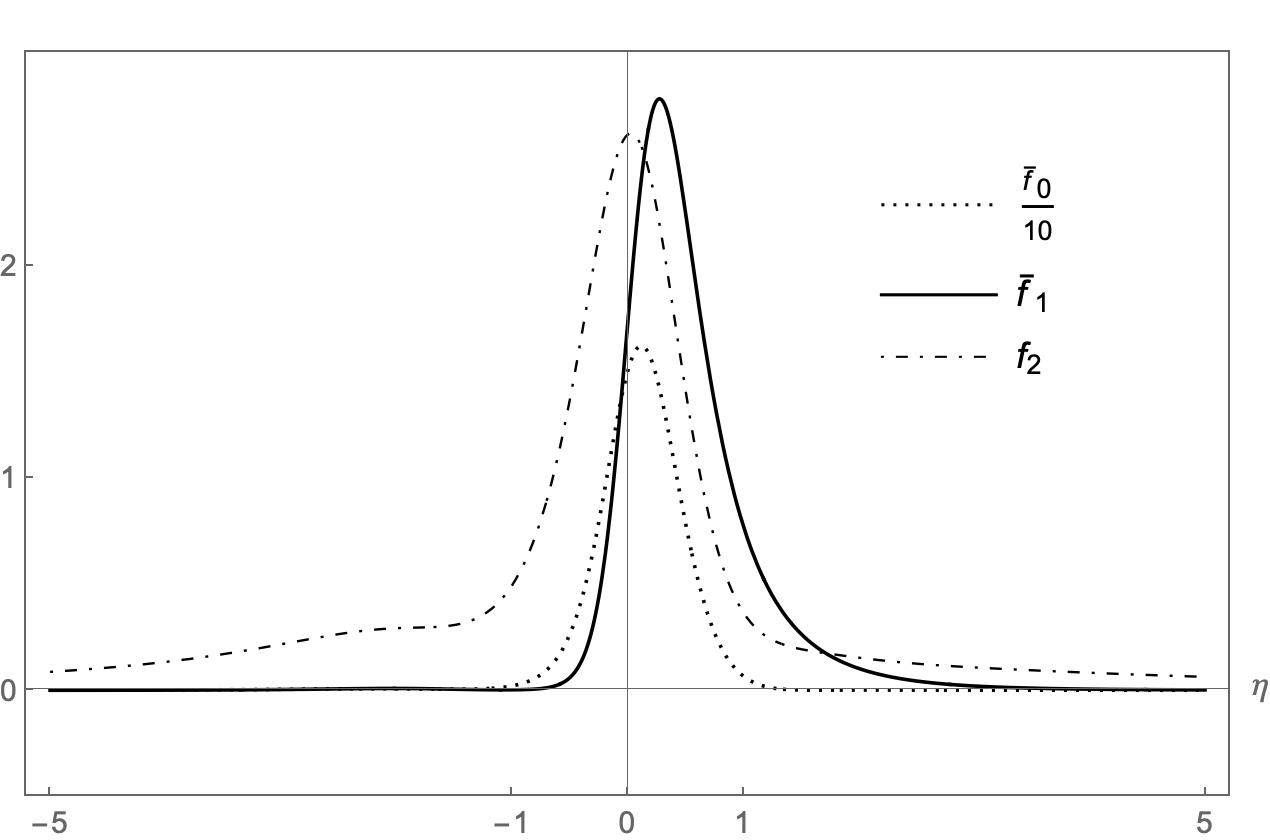}
 \caption{The gradient stability of the graviton at all momenta holds because  $\bar{f}_0>0,\,\bar{f}_1>0 $. The sign of the no-ghost condition for the graviton remains invariant for {\it all momenta} because $ f_2>0$ and $f_3=d_1^2>0$ (That $d_1\neq 0$ holds is clear in the plot of the speed in Figure \ref{fig speeds}, or also in Figure \ref{fig G45}). } \label{fig tensor}
\end{subfigure}%
\begin{subfigure}{0.5\textwidth}
\centering
  \includegraphics[width=1\linewidth]{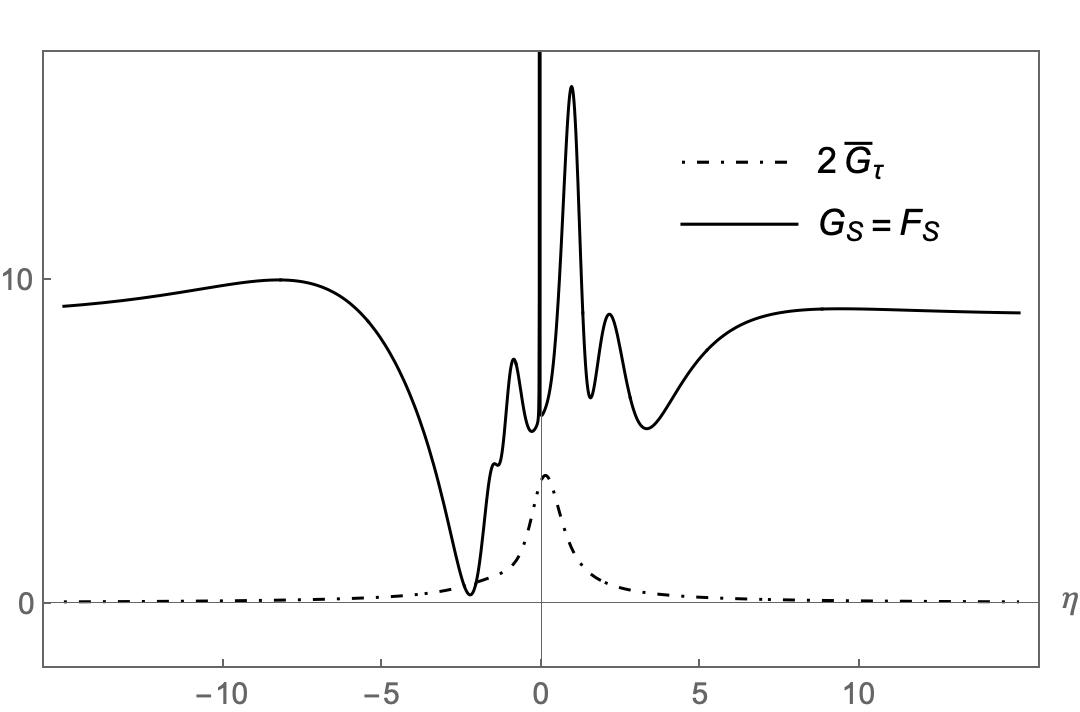}
 \caption{Non ghosty graviton. Stable, non ghosty, luminal scalar mode.} \label{fig scalar}
\end{subfigure}\caption{Stability and no-ghost condition for tensor and scalar modes at all momenta.}
\end{figure}

In our model the background fields are
\begin{eqnarray}
\begin{array}{cccc}
a\,=\, (\tau^2+\eta^2)^{\frac{1}{6}}\,, & H\,=\,\frac{\dot{a}}{a^2}\,=\, \frac{\eta}{3\,(\tau^2+\eta^2)^{\frac{7}{6}}}\,, &\, \phi=\eta\,, &\, x=-\frac{1}{3\left(1+\eta^2\right)^{\frac{1}{6}}} \,,\label{eqn sol backgroundsM}
\end{array}
\end{eqnarray}
and to satisfy the conditions (A)-(C) for the tensor perturbation at all momenta, the following inequalities hold 
\begin{equation}
\begin{array}{cccccc}
\bar{\mathcal{G}}_{\tau}>0 \,,& \bar{f}_0>0\,, & \bar{f}_1>0 \,,& {f}_2>0\,, & f_3=d_1^2>0\,, & c_g^2= \frac{f_1}{f_3}\leq 1\,,
\end{array}\label{eqn minimal highp}
\end{equation}
which we plot in Figures \ref{fig tensor} and \ref{fig speeds}. The fourth inequality in (\ref{eqn minimal highp}) guarantees that in the global factor of the tensor sector the following holds for all $\vec{p}$: $(f_2+ \vec{p} \, {} ^2\,d_1^2)>0$. Hence, the stability and non-ghosty conditions remain the same for all momenta. Similarly, the second inequality is required for the stability of the graviton at all momenta. 

For the (momentum independent) scalar sector the inequalities that must hold are simply those in (\ref{eqn stability conditions}) in the last section, which we plot for our model in Figure \ref{fig scalar}.

\begin{figure}[b]
 \centering
\begin{subfigure}{0.5\textwidth}
\centering
  \includegraphics[width=1\linewidth]{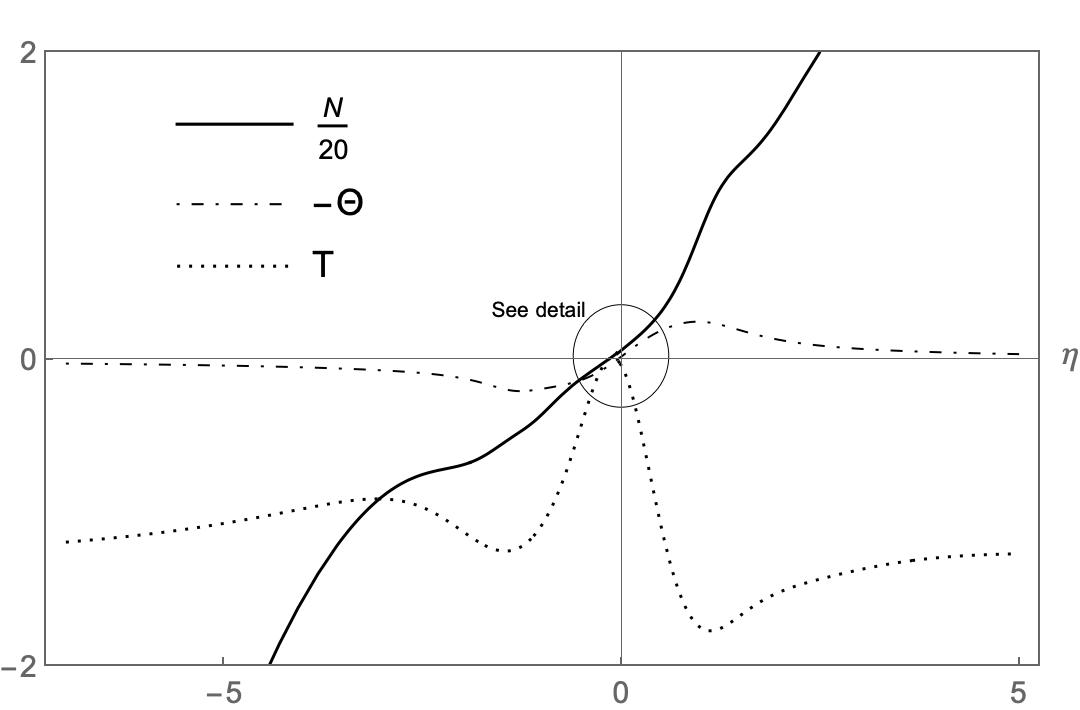}
 \caption{Plot of the function $T$, which we designed to vanish at two points near the bounce phase (We chose $-\Theta=H$). $N$ is monotonous increasing, as required, and vanishes at the zeros of $T$.} \label{}
\end{subfigure}%
\begin{subfigure}{0.5\textwidth}
\centering
  \includegraphics[width=1\linewidth]{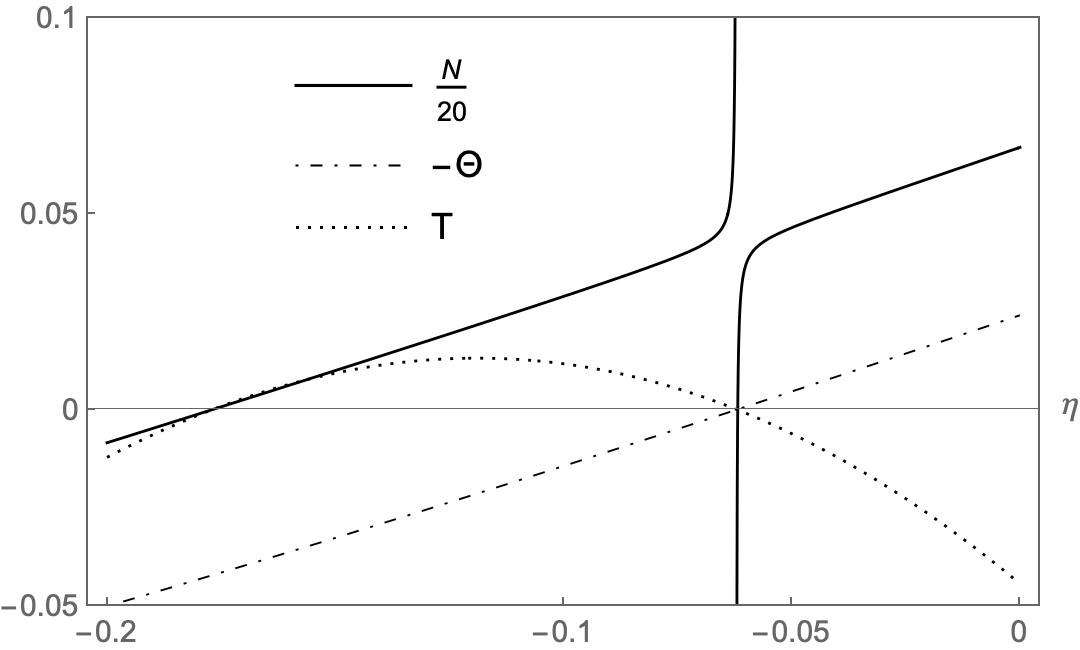}
 \caption{Detail of the vanishing of $T$ and $N$.} \label{}
\end{subfigure}\caption{Avoiding the No-Go}\label{fig T}
\end{figure}

Now, because $N$ must vanish and simultaneously we have no ghosty graviton at low momentum $\mathcal{G}_{\tau}(\vec{p} \, {} ^2=0)= \bar{\mathcal{G}}_{\mathcal{S}}>0$, we have chosen a model where $T$ vanishes, as in the Figure \ref{fig T}, which was the only option left to avoid any pathologies. Importantly, all Euler-Lagrange equations for the background fields $a(\eta),\, \phi,\, x$ (\ref{eqn sol backgroundsM}) are satisfied and the Lagrangian functions are everywhere regular. Indeed, the Ansatz for the Lagrangian functions $G_2$ to $G_5$ is finally solved as in Figures \ref{fig G23}, \ref{fig G45}. In Figure \ref{fig speeds} we also show the Hubble parameter, the speed of the tensor modes and the scalar, and the torsion background.

\section{Conclusions}\label{sec conclusions}

We showed that with the most general Horndeski theories with curvature and torsion it is possible to build classically "healthy" models, namely, with nonsingular, all time stable and subluminal cosmological solutions, at least at linear order in a perturbative expansion.

We first argued how the latter compares to simpler forms of the theory, namely: the historically common form of Horndeski without torsion, and a simplified version of the theory with torsion (namely, without $G_5$). We explained where the usual No-Go arguments break when generalizing from the latter two to the former full Horndeski theory with curvature and torsion.

In brief, we noted that the mathematically and physically unjustified assumption of a torsionless spacetime (or the restriction to up to quartic torsionful theory) leads to accidental relations at linear order, which  restrict the healthiness of the solutions in these simpler variations of the theory, in the form of the "No-Go theorems".

Torsion solves the issues by mere "force", simply because the {\it tensor sector} and the {\it scalar sector} couple {\it in different ways} to  torsion perturbations, thus breaking the usual links between the two sectors, which would otherwise lead to the "No-Go". Namely, despite the fact that the torsion perturbations are non dynamical, these couplings of the metric perturbation and the Horndeski scalar to the Torsion field, greatly modify {\it in diverging ways} the coefficients in the action for the final graviton and the scalar mode. Thus, the stability, sub/ luminality and no ghost criteria restrict very different combinations of the Lagrangian functions for both sectors, allowing to independently satisfy them  without meeting any contradictions at some time in the evolution. 

We also showed that all the healthy models must have a peculiar dispersion relation for the gravitational waves. It opens the question whether there are  phenomenological imprints left by this distinctive graviton. 

We showed a toy model where all the criteria of healthiness can be satisfied. However, the usual construction methods used in the literature ({\it e.g.}  \cite{Ijjas:2016tpn} or \cite{Mironov:2018oec}) prove hard to implement, because the usual criteria for healthiness splits into many conditions that usually cannot be solved algebraically (See Appendix \ref{secapp example}). We presume that obtaining models with physically relevant asymptotics besides the criteria of healthiness is a computational challenge rather than a fundamental obstruction.

\section*{Acknowledgements}

The work of S.M. is partly supported by the grant of the Foundation for the Advancement of Theoretical
Physics and Mathematics “BASIS" and by RFBR grant 21-51-46010.

\section{Appendices}\label{secapp all}

\subsection{Construction of the healthy toy model}\label{secapp example}

In this appendix we show the details of how to construct the model given in section \ref{sec example}.

A reasonable assumption is that taking the all-time nonsingular and stable bouncing solution {\it with a short period of superluminality} that was built in \cite{Mironov:2023wxn}, one can {\it cure the superluminality issue with a contribution from the new $G_5$ part of the action} (\ref{eqn action}). This is our approach below \footnote{This approach is taken by mere simplicity. Namely, close inspection shows that building such a model including $G_5$ with the "reconstruction" method used in \cite{Mironov:2018oec,Mironov:2023wxn} is computationally challenging. The main issue is that simple {\it Ansätze} usually lead to differential equations of the Lagrangian functions $G_A$, which are difficult to solve.}.

\subsubsection{Lagrangian functions $G_3,\, G_4$}

As a first step to a "Healthy model" including $G_5$, let us first build a model with a bouncing solution, but with up to $G_4$. By \cite{Mironov:2023wxn} we know that without $G_5$ the model will suffer at least instabilities or superluminal periods. We will refer to the latter as the "Unhealthy part" of the model. The strategy is then to introduce $G_5$ to remedy these problems.  

For the "Unhealthy part" we will closely follow the example built in \cite{Mironov:2023wxn}. Therefore, for this subsection let us first consider the action (\ref{eqn action}) without $\mathcal{L}_5$. In that case we can also reach the same form of the actions (\ref{eqn Ltensor2}), (\ref{eqn Lscalar}) and we use the same notation for these simplified functions that, let us stress again, {\it only} in this section do not contain $G_5$ nor its derivatives. 

Using in our advantage the generality of Horndeski theory, let us choose the following Ansatz for the general Lagrangian functions $G_3$ and $G_4$:
\begin{eqnarray}
G_3(\phi,X)&=&g_{30}(\phi)\,+\, g_{31}(\phi)\,(X-X_0)\,, \label{eqn G3 ansatz}\\
G_4(\phi,X)&=&g_{40}(\phi)\,+\, g_{41}(\phi)\,(X-X_0)\,, \label{eqn G4 ansatz}
\end{eqnarray}
where we choose $X_0$ as the function $X$ valued on the choice of background fields
\begin{eqnarray}
\begin{array}{cc}
a\,=\, (\tau^2+\eta^2)^{\frac{1}{6}}\,, &\, \phi=\eta \,.\label{eqn sol backgrounds}
\end{array}
\end{eqnarray}
This choice must be such that the Euler-Lagrange equations for the background fields of the full Horndeski-Cartan theory (including $G_5$) are solved. We guarantee it in the sections below by suitably fixing the $g_{30},\, G_2$ Lagrangian functions. 

The parameter $\tau>0$ fixes the maximum of $H$ and the width of the bounce phase (the domain where $\dot{H}(\eta)>0$ around $\eta=0$), as in Figure \ref{fig speeds}. With the solutions (\ref{eqn sol backgrounds}) we have the following function of $\phi$ for the Ansatz:  $X_0=1/(2\, (\tau^2+\phi^2)^{\frac{1}{3}})$ and we choose a short bounce $\tau=1$.

Now, let us start with a similar model as the one built in \cite{Mironov:2023wxn}. Namely, let us solve for $g_{31},\,g_{40},\, g_{41}$ algebraically from the following three equations,

\begin{eqnarray} 
&&\mathcal{F}_{\tau}(g_{40},\,g_{41})=\frac{23}{25}\,, \label{eqn G4 asymptotics practical}\\ 
&&\frac{T(g_{40},\,g_{41})}{\mathcal{F}_{\tau}}=-\frac{7}{5}+\frac{1}{4}\, \text{sech}\left(\frac{29}{50}\frac{\eta}{\tau}\right)^2+\frac{1}{2}\,\text{sech}\left(\frac{2}{5}\left(\frac{\eta}{\tau}+\frac{3}{2}\right)\right)\\
&&\Theta(g_{30},\,g_{31}, g_{40},\,g_{41}) =-H,\,\label{eqn ThetaG234}
\end{eqnarray}
where only for these equations we plug-in the Ansatz 
\begin{eqnarray}
&&x=:-\frac{8\,a^3\,\,H\,X\,G_{4,X}+\,a^2\,\dot{\phi}\,(G_3-2\,G_{4,\phi})}{4\,(G_4+2\,X\,G_{4,X})}\,.\label{eqn solxG234}
\end{eqnarray}
Let us stress that the equation (\ref{eqn solxG234})  is only physically meaningful within the context of {\it up to quartic} Horndeski-Cartan theory (See \cite{Mironov:2023wxn} for more details). Namely, the expression  (\ref{eqn solxG234}) is devoid of meaning for the full Horndeski-Cartan theory with $G_5$ and in this note it is simply an Ansatz for the function $x(\eta)$ that appears in the functions $\mathcal{F}_{\tau}(g_{40},\,g_{41}),\, T(g_{40},\,g_{41}),\, \Theta(g_{30},\,g_{31}, g_{40},\,g_{41})$ in this section. Here, we take this definition for the mere practical purpose of building the "Unhealthy part of the model" in the same way as in \cite{Mironov:2023wxn}. Up to this point {\bf this amounts to nothing more than a choice of a part of the free Lagrangian functions} $G_3,\, G_4$. 

Now, the key issue of solving the Euler-Lagrange equation derived from  the full Horndeski-Cartan theory (with $G_5$) will be resolved later by suitably fixing the free Lagrangian functions $g_{30},\, G_2$.

\subsubsection{Lagrangian functions $G_2,\, G_5$}

Let us take this "Unhealthy part of the model" as a basis. Namely, keeping our choice of $g_{31},\,g_{40},\, g_{41}$ which are functions of time and $g_{30}$, but are independent of $G_5$ and $G_2$, let us now consider the full Horndeski-Cartan action (\ref{eqn action}) with $G_5$, and let us take the Ansatz
\begin{eqnarray}
G_2(\phi,X)&=&g_{20}(\phi)\,+\, g_{21}(\phi)\,(X-X_0) \,+\, g_{22}(\phi)\,(X-X_0)^2\,, \label{eqn G3 ansatz}\\
G_5(\phi,X)&=&g_{50}(\phi)\,+\, g_{51}(\phi)\,(X-X_0) \,+\, g_{52}(\phi)\,(X-X_0)^2\,. \label{eqn G4 ansatz}
\end{eqnarray}
With {\it this particular Ansatz} all equations to satisfy the stability, subluminality and no-ghost criteria become differential equations of the free Lagrangian functions $g_{50},\ g_{51}$. To avoid these issues, we take a practical simple approach and choosing an Ansatz for the functions $g_{50},\,g_{51}$ we show that a model exists with the desired features. Indeed, careful choice of parameters  with a general Ansatz (and potentially fine tuning\footnote{This would most probably be a problem of the Ansatz and our approach. Whether it can be avoided is an open question that we do not address in this note.}) proves that the following choice is enough for our purpose: 
\begin{eqnarray}
g_{50}&=&\frac{1}{100}\left(-94-26\,\text{sech}\left(\frac{1}{2}-\eta\right) +49\,\tanh\left(\frac{3}{2}-\eta\right)-90\,\text{sech}\left(\eta\right)\,\tanh(\eta)\right.\nonumber\\
&-&\left.80\,\text{sech}\left(\frac{7}{10}(\eta+1)\right)\tanh\left(\frac{7}{10}(\eta+1)\right)^2\right) \,,\\
g_{51}&=&\frac{1}{20}\left(95+97\,\text{sech}\left(\frac{37}{25}-\eta\right) +20\,\tanh\left(\eta+1\right)+150\,\text{sech}\left(\frac{3}{5}\left(\eta-\frac{7}{4}\right)\right)\,\tanh\left(\frac{3}{5}\left(\eta-\frac{7}{4}\right)\right)\right.\nonumber\\
&+&\left.20\,\text{sech}\left(\frac{7}{10}\left(\eta+\frac{11}{20}\right)\right)\tanh\left(\frac{7}{10}\left(\eta+ \frac{11}{20}\right)\right)^2\right)\,.
\end{eqnarray}
Now, plugging-in these Lagrangian functions $g_{50},\ g_{51}$ and $g_{31},\,g_{40},\,g_{41}$ of the last section, we can solve for $g_{30},\, g_{52}$ algebraically from the following equations
\begin{eqnarray}
&&\Theta(g_{30},\,g_{52}) =-H+\frac{1}{100}\left(7\,\text{sech}\left(3\left(\eta+\frac{27}{20}\right)\right)-10\,\text{sech}\left(2\,(\eta-1)\right)\right),\,\label{eqn ThetaG2345}\\
&&\mathcal{E}_{{}^0 K_{0ij}}(g_{30})=0\,,\, \label{eqn k0ijG2345}
\end{eqnarray}
where $\mathcal{E}_{f}=0 $, which denotes the Euler-Lagrange equation for the background field $ f$, and $\Theta(g_{30},\,g_{52})$ (which corresponds to the function $\Theta$ obtained from the action (\ref{eqn Lscalar})) are computed from the full Horndeski-Cartan lagrangian (\ref{eqn action}) {\bf with} $G_5$. It is important to note that $\Theta(g_{30},\,g_{52})$ in (\ref{eqn ThetaG2345}) is totally different from the function (\ref{eqn ThetaG234}) $\Theta(g_{30},\,g_{31}, g_{40},\,g_{41})$ of the last section, which was computed without $G_5$ only to choose the Lagrangian functions $g_{31},\,g_{40},\, g_{41}$.

Furthermore, for simplicity -and in order to mimic the $x(\eta)$ profile in \cite{Mironov:2023wxn}- when solving (\ref{eqn ThetaG2345}), (\ref{eqn k0ijG2345}), we have chosen  a model within the full Horndeski-Cartan theories (\ref{eqn action}) whose torsion background is
\begin{eqnarray}
x(\eta)=-\frac{1}{3\left(1+\eta^2\right)^{\frac{1}{6}}}\,.\label{eqn xG2345}
\end{eqnarray}
In other words, by solving the Euler-Lagrange equation (\ref{eqn k0ijG2345}) we have suitably found the Lagrangian function $g_{30}$ that leads to our desired solution (\ref{eqn xG2345}). Thus, the torsion background (\ref{eqn xG2345}) has actual physical meaning because it is used to solve the   equations of motion (\ref{eqn k0ijG2345}) and (\ref{eqn gii practical}) below, in contrast to the function (\ref{eqn solxG234}) that we took as Ansatz for mere practical purposes in three {\it ad-hoc} equations in the last section just to select the Lagrangian functions $g_{31},\,g_{40},\,g_{41}$.

The approach in this section of fixing $x(\eta)$ and finding $g_{30}$ from the Euler-Lagrange equation is simpler because $\mathcal{E}_{{}^0 K_{0ij}}(g_{30})$ is quadratic in the former and linear in the latter.

Finally, to make our choices of background fields $a(\eta),\,\phi(\eta),\, x(\eta)$ (\ref{eqn sol backgrounds}) and (\ref{eqn xG2345}) consistent, we must satisfy the remaining Euler-Lagrange equations. We do this by choosing a model within the general action (\ref{eqn action}) (with $G_5$) with $g_{20},\,g_{21},\,g_{22}$ solving algebraically the three independent equations
\begin{eqnarray}
\begin{array}{ccc}
\mathcal{G}_{\mathcal{S}}=\mathcal{F}_{\mathcal{S}} \label{eqn Gs practical}\,,& \mathcal{E}_{g_{00}}=0 \label{eqn g00 practical}\,, & \mathcal{E}_{g_{11}}= \mathcal{E}_{g_{22}}= \mathcal{E}_{g_{33}}=0 \label{eqn gii practical}\,,
\end{array}
\end{eqnarray}
of which the first equation implies a model with a {\it scalar mode that propagates exactly at the speed of light}.

\begin{figure}[t]
 \centering
\begin{subfigure}{0.5\textwidth}
\centering
  \includegraphics[width=1\linewidth]{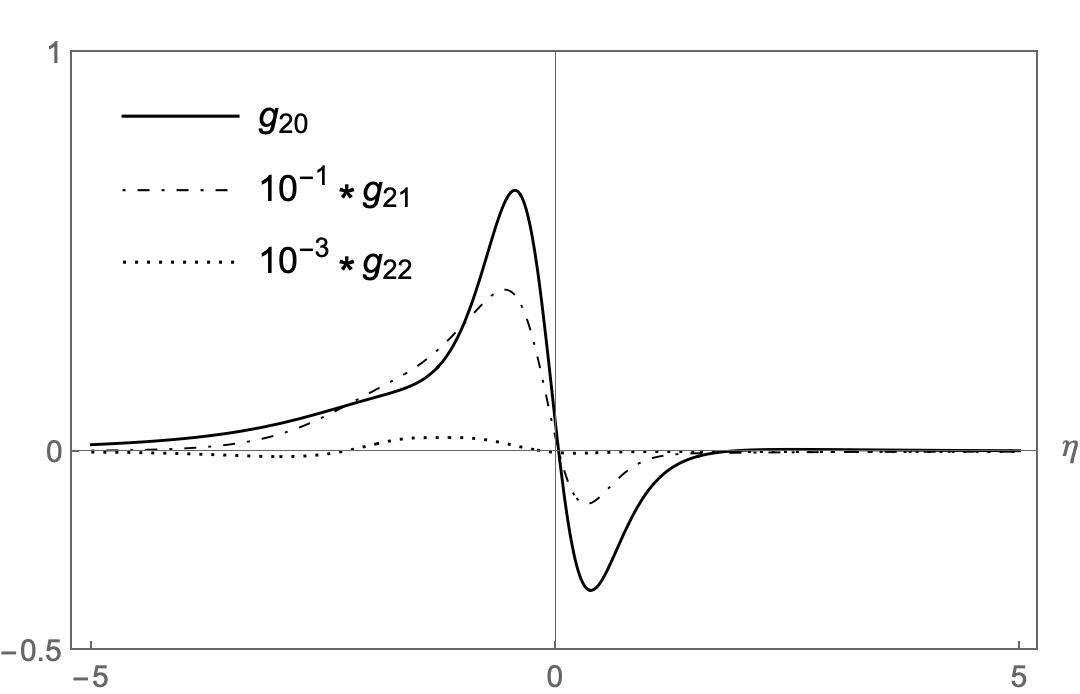}
 \caption{Plot of $g_{20},\, g_{21},\, g_{22}$ which we chose to satisfy two of the equations of motion for the background fields, and the luminality of the scalar.} \label{}
\end{subfigure}%
\begin{subfigure}{0.5\textwidth}
\centering
  \includegraphics[width=1\linewidth]{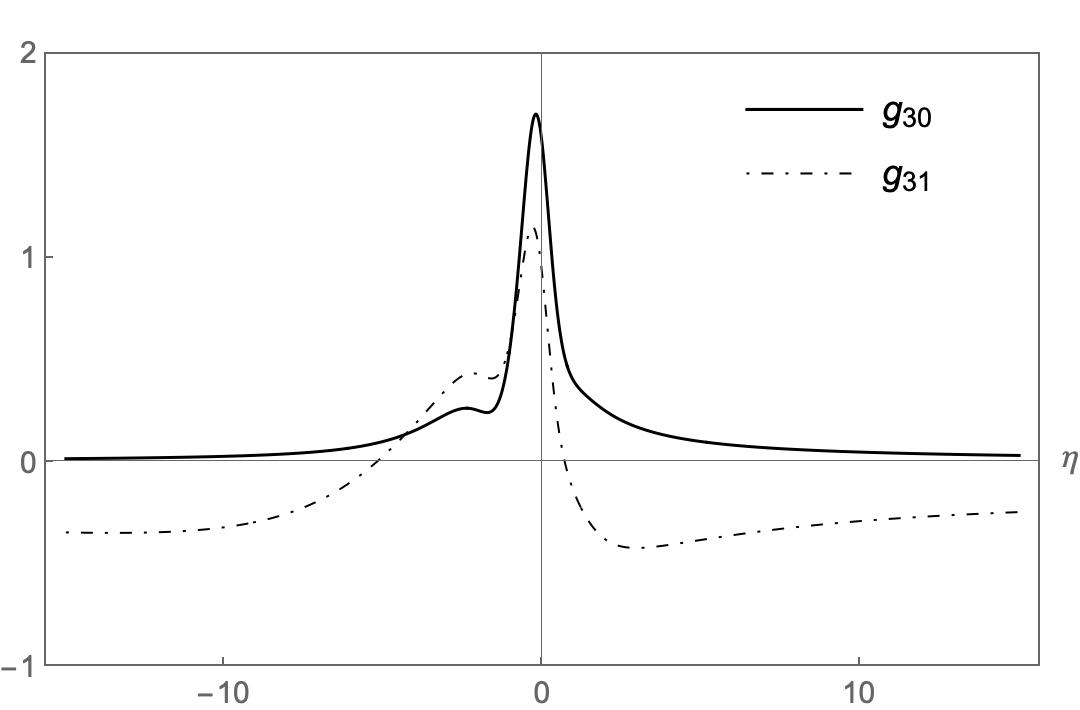}
 \caption{Plot of $g_{30},\, g_{31}$ which we chose to satisfy the third independent equation of motion of background fields, and to fix the function $\Theta$} \label{}
\end{subfigure}\caption{Everywhere regular Lagrangian functions $G_2,\, G_3$}\label{fig G23}
\end{figure}

\begin{figure}[t]
 \centering
\begin{subfigure}{0.5\textwidth}
\centering
  \includegraphics[width=1\linewidth]{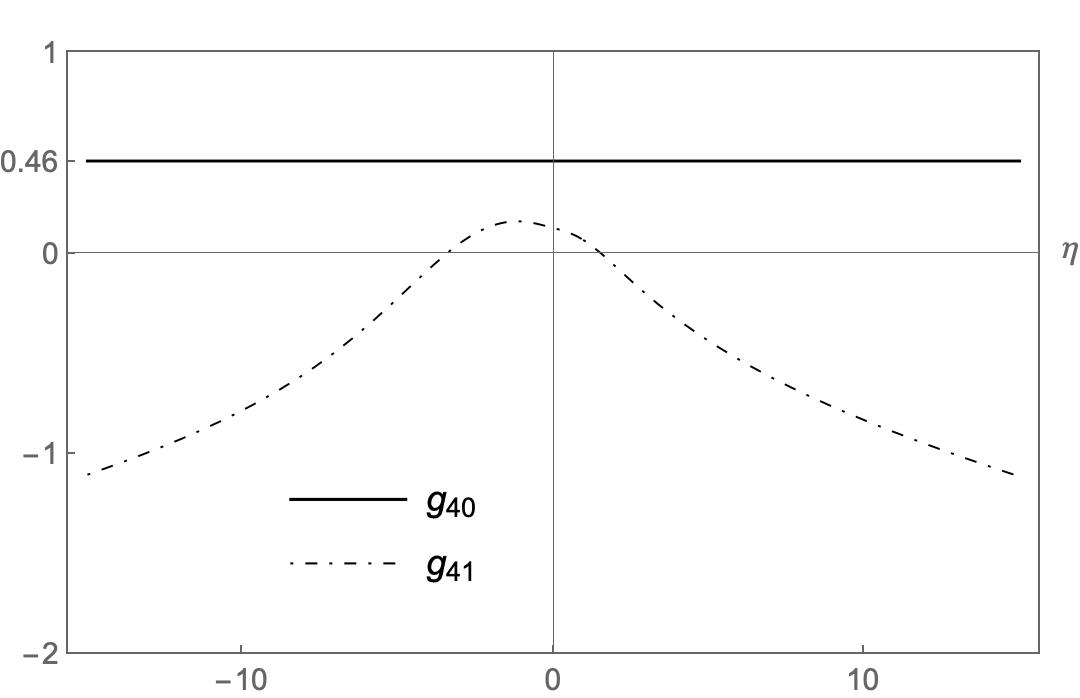}
 \caption{Plot of $g_{40},\, g_{41}$} \label{}
\end{subfigure}%
\begin{subfigure}{0.5\textwidth}
\centering
  \includegraphics[width=1\linewidth]{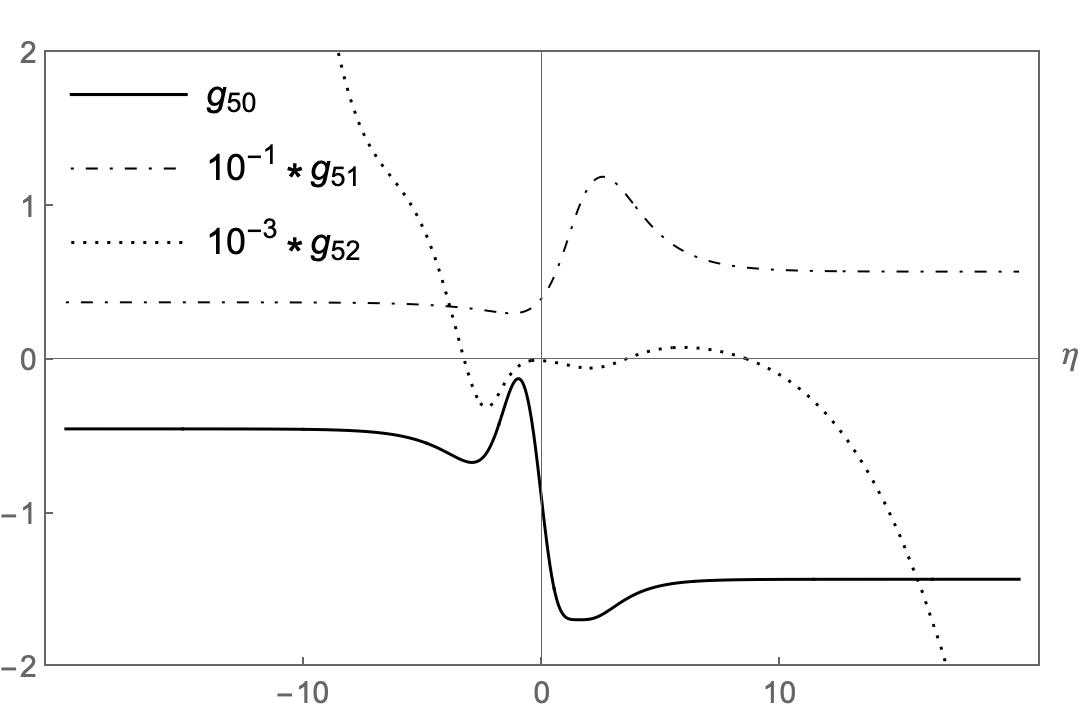}
 \caption{Plot of $g_{50},\, g_{51}$, which we carefully chose starting with an {\it ad-hoc} Ansatz and with the sole criteria of obtaining all-time stability and subluminality of the bounce. } \label{}
\end{subfigure}\caption{Everywhere regular Lagrangian functions $G_4,\, G_5$. Let us notice that $f_3\propto d_1\propto G_5=g_{50}$ on-shell, which does not vanish. Hence, the speed of gravitational waves $\propto f_3^{-1}$ in Eq. (\ref{eqn cg2}) is well defined.}\label{fig G45}
\end{figure}

The solution for the Lagrangian functions are nonsingular everywhere. They are plotted in Figures \ref{fig G23} and \ref{fig G45}.

\subsection{Notation for torsion and its  decomposition into irreducible components}\label{secapp notation}

We consider torsion in the second order (metric formalism). We closely follow the notation in \cite{Mironov:2023kzt,Mironov:2023wxn}. We give the necessary details below for completeness. As before, let us denote the torsion tensor as the difference between the non-symmetric connection:
\begin{eqnarray}
T^{\rho}{}_{\mu\nu}=\tilde{\Gamma}^{\rho}_{\mu\nu}-\tilde{\Gamma}^{\rho}_{\nu\mu}\,,
\end{eqnarray}
and for convenience we also introduce the contortion tensor:
\begin{eqnarray}
K^{\rho}{}_{\mu\nu}&=&-\frac{1}{2}\left(T_{\nu}{}^{\rho}{}_{\mu}+T_{\mu}{}^{\rho}{}_{\nu}+T^{\rho}{}_{\mu\nu}\right)\,,\label{eqn ktrelation}
\end{eqnarray}
and let us notice the antisymmetry $T^{\rho}{}_{\mu\nu}= -T^{\rho}{}_{\nu\mu}$, $K_{\mu\nu\sigma}= -K_{\sigma\nu\mu}$. It is easy to see that we can express torsionful quantities in terms of torsionless covariant derivatives ($\nabla$) associated with the Christoffel connection (${\Gamma}^{\rho}_{\mu\nu}$), plus contortion, as follows: 
\begin{eqnarray}
\tilde{\Gamma}^{\rho}_{\mu\nu}&=&\Gamma^{\rho}_{\mu\nu}-K^{\rho}{}_{\mu\nu}\,,
\end{eqnarray}
\begin{eqnarray}
\tilde{\nabla}_\mu V^\nu=\partial_\mu V^\nu+\tilde{\Gamma}^{\nu}_{\mu\lambda}V^{\lambda}\,\, \label{eqn torsionless covd}
\end{eqnarray}
\begin{eqnarray}
\tilde{\nabla}_\mu V^\nu=\nabla_\mu V^\nu-K^{\nu}{}_{\mu\lambda}V^{\lambda}\,,\label{eqn covds}
\end{eqnarray}
where we draw attention to our convention to sum over the second lower index of the torsionful connection in Eq (\ref{eqn torsionless covd}). Thus, we can explicitly write the action (\ref{eqn action}) in terms of contortion, the metric and the Horndeski scalar using Eq. (\ref{eqn covds}), such as in \cite{Mironov:2023kzt}. For instance, we can rewrite the Ricci tensor computed with the nonsymmetric connection, $\tilde{R}$, in terms of the Ricci tensor computed with the Christoffel connection, $R$, as
\begin{equation}
\tilde{R}_{\mu\nu}=R_{\mu\nu}+\nabla_\nu K^{\rho}{}_{\rho\mu}-\nabla_\rho K^{\rho}{}_{\nu\mu}+K^{\rho}{}_{\rho\sigma}\,K^{\sigma}{}_{\nu\mu}-K^{\rho}{}_{\nu\sigma}\,K^{\sigma}{}_{\rho\mu}\,,\label{eqn riccit}
\end{equation}
and similarly for the Einstein tensor. Thus, the latter are not symmetric \cite{Hehl:1976kj} and they introduce far from trivial $\mathcal{O}(K^2,\,K^3)$ contributions in the first term of the action (\ref{eqn L5}), $\,G_5 \,\tilde{G}^{\mu\nu}\, \tilde{\nabla}_\mu \tilde{\nabla}_\nu\phi$, which, as dicussed before, is responsible for the momentum dependent dispersion relation of the graviton (\ref{eqn cgw}). More precisely, the $p^2\, T^{\scalebox{0.5}{(1)}}_{ij}\, T^{\scalebox{0.5}{(2)}}_{ij}   $, which is essential to   the stark modification of the graviton in comparison to other forms of Horndeski theory, stems from the third term of the Ricci tensor in Eq.  (\ref{eqn riccit}) plugging in the first term of the action (\ref{eqn L5}):  $\,G_5 \,(\nabla_\rho K^{\rho\nu\mu})\, \tilde{\nabla}_\mu \tilde{\nabla}_\nu\phi$ . This  turns out to be essential to allow globally healthy Horndeski gravities.

\subsubsection{Linearization}\label{secapp decomposition and lin}

For the perturbative expansion at linear order about a spatially flat FLRW background let us decompose the perturbations into irreducible components under small rotation group. We show all details below for completeness. The perturbed metric is denoted as
\begin{eqnarray}
\textrm{d}s^2=\left(\eta_{\mu\nu}+\delta g_{\mu\nu}\right)\textrm{d}x^\mu\, \textrm{d}x^\nu
\end{eqnarray}
where 
\begin{eqnarray}
\eta_{\mu\nu}\textrm{d}x^\mu\, \textrm{d}x^\nu= a^2(\eta)\left(-\textrm{d}\eta^2+\delta_{ij}\, \textrm{d}x^i \,\textrm{d}x^j \right)\label{eqn backgroundmetric}
\end{eqnarray}
and $\eta$ is conformal time. The metric perturbation is
\begin{eqnarray}
\delta g_{\mu\nu}\,\textrm{d}x^\mu\, \textrm{d}x^\nu
=a^2(\eta)\left(-2\,\alpha\,\textrm{d}\eta^2+2\left(\partial_i B+S_i\right) \textrm{d}\eta \, \textrm{d}x^i+\left(-2\,\psi\, \delta_{ij}+2\,\partial_i\partial_j E+\partial_i F_j+\partial_j F_i+2\,h_{ij}\right) \textrm{d}x^i \, \textrm{d}x^j \right)\,,
\end{eqnarray}
with $\alpha,\, B,\, \psi,\, E$  scalar perturbations, $S_i,\, F_i$  transverse vector perturbations, and $h_{ij}$, a symmetric, traceless and transverse tensor perturbation.\bigskip

For contortion, antisymmetric in the first and third indices, there are $24$ independent components. They are written as eight scalars denoted as $C^{\scalebox{0.5}{(n)}} $ with $n=1, \dots , 8$, six (two-component) transverse vectors denoted as $V^{\scalebox{0.5}{(m)}}_i$ with $m=1,\dots, 6$ and two (two-component) traceless, symmetric, transverse tensors $T^{\scalebox{0.5}{(1)}}_{ij},\, T^{\scalebox{0.5}{(2)}}_{ij}$. 

Explicitly, the decomposition of contortion perturbation reads, for the scalar sector
\begin{eqnarray}
\delta K^{\text{scalar}}_{i00}&=&\partial_i C^{\scalebox{0.5}{(1)}} \nonumber \\
\delta K^{\text{scalar}}_{ij0}&=& \partial_i \partial_j C^{\scalebox{0.5}{(2)}} +\delta_{ij} C^{\scalebox{0.5}{(3)}} +\epsilon_{ijk} \partial_k C^{\scalebox{0.5}{(4)}} \nonumber \\
\delta K^{\text{scalar}}_{i0k}&=&\epsilon_{ikj} \partial_j C^{\scalebox{0.5}{(5)}} \nonumber \\
\delta K^{\text{scalar}}_{ijk}&=&\left(\delta_{ij} \partial_k-\delta_{kj} \partial_i\right) C^{\scalebox{0.5}{(6)}} +\epsilon_{ikl} \partial_l \partial_j C^{\scalebox{0.5}{(7)}} +\left(\epsilon_{ijl} \partial_l \partial_k-\epsilon_{kjl} \partial_l \partial_i\right) C^{\scalebox{0.5}{(8)}} \label{eqn kspert}\,,
\end{eqnarray}
for the vector sector
\begin{eqnarray}
\delta K^{\text{vector}}_{i00}&=& V^{\scalebox{0.5}{(1)}}_i\nonumber\\
\delta K^{\text{vector}}_{ij0}&=& \partial_i V^{\scalebox{0.5}{(2)}}_j+ \partial_j V^{\scalebox{0.5}{(3)}}_i\nonumber\\
\delta K^{\text{vector}}_{i0k}&=& \partial_i V^{\scalebox{0.5}{(4)}}_k- \partial_k V^{\scalebox{0.5}{(4)}}_i\nonumber\\
\delta K^{\text{vector}}_{ijk}&=&\delta_{ij} V^{\scalebox{0.5}{(5)}}_k-\delta_{kj} V^{\scalebox{0.5}{(5)}}_i + \partial_j \partial_i V^{\scalebox{0.5}{(6)}}_k-\partial_j \partial_k V^{\scalebox{0.5}{(6)}}_i \label{eqn kvpert}\,,
\end{eqnarray}
and for the tensor sector
\begin{eqnarray}
\delta K^{\text{tensor}}_{ij0}&=& T^{\scalebox{0.5}{(1)}}_{ij}\nonumber\\
\delta K^{\text{tensor}}_{ijk}&=& \partial_i T^{\scalebox{0.5}{(2)}}_{jk}-\partial_k T^{\scalebox{0.5}{(2)}}_{ji} \label{eqn ktpert}\,,
\end{eqnarray}
Thus, the components of contortion perturbation are
\begin{eqnarray}
\delta K_{i\mu\nu}= \delta K^{\text{scalar}}_{i\mu\nu}+ \delta K^{\text{vector}}_{i\mu\nu}+ \delta K^{\text{tensor}}_{i\mu\nu}\,.
\end{eqnarray}
The non-vanishing components of the background contortion tensor on a homogeneous and isotropic background spacetime are 
\begin{eqnarray}
{}^{0}K_{0jk}=x(\eta)\delta_{jk}\nonumber \\
{}^{0}K_{ijk}=y(\eta)\epsilon_{ijk}\,,
\end{eqnarray}
thus the contortion tensor with all indices down is written as:
\begin{eqnarray}
K_{\mu\nu\sigma}= {}^{0}K_{\mu\nu\sigma}+ \delta K_{\mu\nu\sigma}
\end{eqnarray}
Finally, the scalar field is written in terms of a time dependent part  $\phi(\eta)$ plus a  spacetime dependent perturbation $\Pi$. The distinction between the background field $\phi(\eta)$ and the spacetime dependent field $\phi(x)$ will be clear from the context.

\subsection{Equations of motion for the background fields}\label{secapp equations backgrounds}

From (\ref{eqn action}) we can compute the action for the background fields $a,\,\phi,\,x,\,y$. The Euler-Lagrange equations $\mathcal{E}_{f}=\frac{\partial \mathcal{L}}{\partial f}=0$ for $f$ one of the following components $g_{00},\,g_{ij},\,K_{ij0},\,K_{ijk},\phi$ are of up to second order in the background fields. In particular, 

\begin{dmath}\mathcal{E}_{K_{ijk}}=- y\, \frac{2 \,{a}^4 \,{G_{4}}  +  \,{G_{5}} \left(\,{a}^2  \,\ddot{\phi} +  \,{x}  \,\dot{\phi}\right)}{\,{a}^{10}} \,,\label{eqn y}\end{dmath}
leads to two branches of background solutions: namely, either the torsion background $y$ vanishes or not. Throughout this note we have assumed the simpler, former branch 
\begin{equation}
y(\eta)\equiv 0\,.\label{eqn soly}
\end{equation} 
Let us notice that if there is no $G_5$ in the action (up to quartic Horndeski-Cartan), Eq. (\ref{eqn soly}) would be the only possible solution to Eq. (\ref{eqn y}). Thus, the branch (\ref{eqn soly}) is a natural continuation to previous results in up to quartic Horndeski-Cartan theory in \cite{Mironov:2023kzt,Mironov:2023wxn,valenciavillegas}.

With (\ref{eqn soly}) the remaining equations can be written with
\begin{dmath}\mathcal{E}_{g_{00}}=\frac{1}{2 \,{a}^{12}}\Big(\,{a}^{10} (- \,{G_{2}} - 6 \,{G_{4}} \,{H}^2 + (2 \,{G_{2,X}} - 2 \,{G_{3,{\phi}}} + (24 \,{G_{4,X}} - 18 \,{G_{5,{\phi}}}) \,{H}^2) \,{X} + (24 \,{G_{4,XX}} - 12 \,{G_{5,{\phi}X}}) \,{H}^2 \,{X}^2) + \,{a}^7 (\,{H} (72 \,{G_{4,X}} \,{x} - 36 \,{G_{5,{\phi}}} \,{x}) \,{X} + \,{H} (48 \,{G_{4,XX}} \,{x} - 24 \,{G_{5,{\phi}X}} \,{x}) \,{X}^2) + \,{a}^4 (6 \,{G_{4}} \,{x}^2 + (48 \,{G_{4,X}} \,{x}^2 - 18 \,{G_{5,{\phi}}} \,{x}^2) \,{X} + (24 \,{G_{4,XX}} \,{x}^2 - 12 \,{G_{5,{\phi}X}} \,{x}^2) \,{X}^2) + \,{a}^9 (-6 \,{G_{4,{\phi}}} \,{H} + ((6 \,{G_{3,X}} - 12 \,{G_{4,{\phi}X}}) \,{H} + 10 \,{G_{5,X}} \,{H}^3) \,{X} + 4 \,{G_{5,XX}} \,{H}^3 \,{X}^2) \,\dot{\phi} + \,{a}^6 (3 \,{G_{3}} \,{x} - 6 \,{G_{4,{\phi}}} \,{x} + 9 \,{G_{5}} \,{H}^2 \,{x} + (6 \,{G_{3,X}} \,{x} - 12 \,{G_{4,{\phi}X}} \,{x} + 36 \,{G_{5,X}} \,{H}^2 \,{x}) \,{X} + 12 \,{G_{5,XX}} \,{H}^2 \,{x} \,{X}^2) \,\dot{\phi} + \,{a}^3 (18 \,{G_{5}} \,{H} \,{x}^2 + 42 \,{G_{5,X}} \,{H} \,{x}^2 \,{X} + 12 \,{G_{5,XX}} \,{H} \,{x}^2 \,{X}^2) \,\dot{\phi} + (9 \,{G_{5}} \,{x}^3 + 16 \,{G_{5,X}} \,{x}^3 \,{X} + 4 \,{G_{5,XX}} \,{x}^3 \,{X}^2) \,\dot{\phi}\Big) \,,\end{dmath}

\begin{dmath}\mathcal{E}_{g_{ij}}=\frac{\delta_{ij}}{2 \,{a}^{12}}\,\Big(\,{a}^{10} (\,{G_{2}} + 6 \,{G_{4}} \,{H}^2 + (-2 \,{G_{3,{\phi}}} + 4 \,{G_{4,{\phi}{\phi}}}) \,{X}) + (-3 \,{G_{5}} \,{x}^3 - 2 \,{G_{5,X}} \,{x}^3 \,{X}) \,\dot{\phi} + \,{a}^6 (\,{G_{3}} \,{x} - 2 \,{G_{4,{\phi}}} \,{x} + 3 \,{G_{5}} \,{H}^2 \,{x} + (-8 \,{G_{4,{\phi}X}} \,{x} + 4 \,{G_{5,{\phi}{\phi}}} \,{x} + 18 \,{G_{5,X}} \,{H}^2 \,{x}) \,{X} + 8 \,{G_{5,XX}} \,{H}^2 \,{x} \,{X}^2) \,\dot{\phi} + \,{a}^3 (6 \,{G_{5}} \,{H} \,{x}^2 + 14 \,{G_{5,X}} \,{H} \,{x}^2 \,{X} + 4 \,{G_{5,XX}} \,{H} \,{x}^2 \,{X}^2) \,\dot{\phi} + \,{a}^9 (4 \,{G_{4}} \,\dot{H} + (2 \,{G_{4,{\phi}}} \,{H} + ((2 \,{G_{3,X}} - 12 \,{G_{4,{\phi}X}} + 4 \,{G_{5,{\phi}{\phi}}}) \,{H} + 2 \,{G_{5,X}} \,{H}^3) \,{X} + 4 \,{G_{5,XX}} \,{H}^3 \,{X}^2) \,\dot{\phi}) + \,{a}^8 (2 \,{G_{4,{\phi}}} \,\ddot{\phi} - 4 \,{G_{5,XX}} \,{H}^2 \,{X}^2 \,\ddot{\phi} + \,{X} (-2 \,{G_{3,X}} \,\ddot{\phi} + 4 \,{G_{4,{\phi}X}} \,\ddot{\phi} - 6 \,{G_{5,X}} \,{H}^2 \,\ddot{\phi}) - 4 \,{G_{5,X}} \,{H} \,{X} \,\dot{H} \,\dot{\phi} + ((-2 \,{G_{4,X}} -  \,{G_{5,{\phi}}}) \,{H}^2 + (8 \,{G_{4,XX}} - 6 \,{G_{5,{\phi}X}}) \,{H}^2 \,{X}) \,\dot{\phi}^2) + \,{a}^7 ((\,{H} (-4 \,{G_{4,X}} \,\ddot{\phi} + 4 \,{G_{5,{\phi}}} \,\ddot{\phi}) + \,{H} \,{X} (-8 \,{G_{4,XX}} \,\ddot{\phi} + 4 \,{G_{5,{\phi}X}} \,\ddot{\phi})) \,\dot{\phi} + (-4 \,{G_{4,X}} + 2 \,{G_{5,{\phi}}}) \,\dot{H} \,\dot{\phi}^2) + \,{a}^4 (-2 \,{G_{4}} \,{x}^2 + (-8 \,{G_{4,X}} \,{x} \,\ddot{\phi} + 4 \,{G_{5,{\phi}}} \,{x} \,\ddot{\phi} + \,{X} (-8 \,{G_{4,XX}} \,{x} \,\ddot{\phi} + 4 \,{G_{5,{\phi}X}} \,{x} \,\ddot{\phi})) \,\dot{\phi} + \,\dot{\phi}^2 (-4 \,{G_{4,X}} \,\dot{x} + 2 \,{G_{5,{\phi}}} \,\dot{x})) + \,{a}^2 (-2 \,{G_{5}} \,{x}^2 \,\ddot{\phi} - 10 \,{G_{5,X}} \,{x}^2 \,{X} \,\ddot{\phi} - 4 \,{G_{5,XX}} \,{x}^2 \,{X}^2 \,\ddot{\phi} + (-2 \,{G_{4,X}} \,{x}^2 -  \,{G_{5,{\phi}}} \,{x}^2 - 2 \,{G_{5,{\phi}X}} \,{x}^2 \,{X}) \,\dot{\phi}^2 + \,\dot{\phi} (-4 \,{G_{5}} \,{x} \,\dot{x} - 4 \,{G_{5,X}} \,{x} \,{X} \,\dot{x})) + \,{a}^5 (-2 \,{G_{5}} \,{H} \,{x} \,\ddot{\phi} - 16 \,{G_{5,X}} \,{H} \,{x} \,{X} \,\ddot{\phi} - 8 \,{G_{5,XX}} \,{H} \,{x} \,{X}^2 \,\ddot{\phi} + (\,{H} (12 \,{G_{4,X}} \,{x} - 8 \,{G_{5,{\phi}}} \,{x}) + \,{H} (8 \,{G_{4,XX}} \,{x} - 8 \,{G_{5,{\phi}X}} \,{x}) \,{X}) \,\dot{\phi}^2 + \,\dot{\phi} (-2 \,{G_{5}} \,{x} \,\dot{H} - 2 \,{G_{5}} \,{H} \,\dot{x} + \,{X} (-4 \,{G_{5,X}} \,{x} \,\dot{H} - 4 \,{G_{5,X}} \,{H} \,\dot{x})))\Big) \,,\end{dmath}

\begin{dmath}\mathcal{E}_{K_{ij0}}=\frac{\delta_{ij}}{2 \,{a}^{10}}\Big(\,{a}^7 (-8 \,{G_{4,X}} + 4 \,{G_{5,{\phi}}}) \,{H} \,{X} + \,{a}^4 (-4 \,{G_{4}} \,{x} + (-8 \,{G_{4,X}} \,{x} + 4 \,{G_{5,{\phi}}} \,{x}) \,{X}) + \,{a}^6 (- \,{G_{3}} + 2 \,{G_{4,{\phi}}} -  \,{G_{5}} \,{H}^2 - 2 \,{G_{5,X}} \,{H}^2 \,{X}) \,\dot{\phi} + \,{a}^3 (-4 \,{G_{5}} \,{H} \,{x} - 4 \,{G_{5,X}} \,{H} \,{x} \,{X}) \,\dot{\phi} + (-3 \,{G_{5}} \,{x}^2 - 2 \,{G_{5,X}} \,{x}^2 \,{X}) \,\dot{\phi}\Big) \,,\end{dmath}
and, due to gauge redundancy, the equation for $\phi$ is trivially satisfied by the latter and their time derivatives. Indeed,  
\begin{equation}
\mathcal{E}_{\phi}=a^2\,\left(\dot{\mathcal{E}}_{g_{00}}+a\,\left(5\,\mathcal{E}_{g_{00}}+3\,\mathcal{E}_{g_{ii}}\right)\,H\right)+3\,x\,\left(\dot{\mathcal{E}}_{K_{ii0}}+4\,a\,H\,\mathcal{E}_{K_{ii0}}\right)\,,\label{eqn genphi}
\end{equation}
where repeated spatial indices are not summed in (\ref{eqn genphi}).

\subsection{Quadratic action for the tensor sector}\label{secapp coefficients tensor modes}

From (\ref{eqn action}) we can write the quadratic action for the tensor modes in the form (\ref{eqn Ltensor1}). The coefficients $b_A,\, c_A,\, d_A$ are functions of time {\it only}. They are given by the following expressions:

\begin{dmath}b_{1}=\,{a}^2 (\,{G_{4}} + (-2 \,{G_{4,X}} + \,{G_{5,{\phi}}}) \,{X}) -  \,{a} \,{G_{5,X}} \,{H} \,{X} \,\dot{\phi} -  \frac{\,{x} (\,{G_{5}} + 2 \,{G_{5,X}} \,{X}) \,\dot{\phi}}{2 \,{a}^2} \,,\end{dmath}\begin{dmath}b_{2}=- \frac{2 \,{a}^4 (\,{G_{4}} -  \,{G_{5,{\phi}}} \,{X}) + \,{a} \,{G_{5,X}} \,{H} \,\dot{\phi}^3 -  \,\dot{\phi} (3 \,{G_{5}} \,{x} + \,{G_{5,X}} \,\ddot{\phi} \,\dot{\phi})}{2 \,{a}^2} \,,\end{dmath}\begin{dmath}b_{3}=- \frac{2}{\,{a}^6}\, \Big(\,{x}^3 (3 \,{G_{5}} + 2 \,{G_{5,X}} \,{X}) \,\dot{\phi} - 4 \,{a}^3 \,{H} \,{x}^2 (\,{G_{5}} + \,{X} (3 \,{G_{5,X}} + \,{G_{5,XX}} \,{X})) \,\dot{\phi} -  \,{a}^6 \,{x} (2 (-2 \,{G_{4,{\phi}X}} + \,{G_{5,{\phi}{\phi}}}) \,{X} + \,{H}^2 (\,{G_{5}} + 4 \,{X} (2 \,{G_{5,X}} + \,{G_{5,XX}} \,{X}))) \,\dot{\phi} + 2 \,{a}^2 \,{x} (\,{x} (\,{G_{5}} + \,{X} (5 \,{G_{5,X}} + 2 \,{G_{5,XX}} \,{X})) \,\ddot{\phi} + \,{x} (2 \,{G_{4,X}} + \,{G_{5,{\phi}X}} \,{X}) \,\dot{\phi}^2 + 2 (\,{G_{5}} + \,{G_{5,X}} \,{X}) \,\dot{\phi} \,\dot{x}) -  \,{a}^5 (\,{x} (- \,{H} (\,{G_{5}} + 4 \,{X} (2 \,{G_{5,X}} + \,{G_{5,XX}} \,{X})) \,\ddot{\phi} -  (\,{G_{5}} + 2 \,{G_{5,X}} \,{X}) \,\dot{H} \,\dot{\phi} + \,{H} (4 \,{G_{4,X}} - 3 \,{G_{5,{\phi}}} + 4 \,{G_{4,XX}} \,{X} - 4 \,{G_{5,{\phi}X}} \,{X}) \,\dot{\phi}^2) -  \,{H} (\,{G_{5}} + 2 \,{G_{5,X}} \,{X}) \,\dot{\phi} \,\dot{x}) -  \,{a}^4 (2 \,{x} (-2 \,{G_{4}} \,{x} + (-2 \,{G_{4,X}} + \,{G_{5,{\phi}}} - 2 \,{G_{4,XX}} \,{X} + \,{G_{5,{\phi}X}} \,{X}) \,\ddot{\phi} \,\dot{\phi}) + (-2 \,{G_{4,X}} + \,{G_{5,{\phi}}}) \,\dot{\phi}^2 \,\dot{x})\Big) \,,\end{dmath}

\begin{dmath}c_{1}=\frac{2 \,{a}^4 \,{G_{4}} + \,{a}^2 \,{G_{5}} \,\ddot{\phi} + \,{G_{5}} \,{x} \,\dot{\phi}}{2 \,{a}^6} \,,\end{dmath}\begin{dmath}c_{2}=- \frac{2 \,{x} (2 \,{a}^4 (\,{G_{4}} + 2 \,{G_{4,X}} \,{X} -  \,{G_{5,{\phi}}} \,{X}) + 2 \,{a}^3 \,{H} (\,{G_{5}} + \,{G_{5,X}} \,{X}) \,\dot{\phi} + \,{x} (3 \,{G_{5}} + 2 \,{G_{5,X}} \,{X}) \,\dot{\phi})}{\,{a}^6} \,,\end{dmath}\begin{dmath}c_{3}=\frac{\,\dot{\phi} (2 \,{x} (\,{G_{5}} + \,{G_{5,X}} \,{X}) + \,{a}^3 \,{H} (\,{G_{5}} + 2 \,{G_{5,X}} \,{X}) + \,{a}^2 (2 \,{G_{4,X}} -  \,{G_{5,{\phi}}}) \,\dot{\phi})}{\,{a}^4} \,,\end{dmath}

\begin{dmath}c_{4}=\frac{1}{4\,x}\,c_2 \,,\end{dmath}

\begin{eqnarray}d_{1}=\frac{ \,{G_{5}} \,\dot{\phi}}{\,{a}^4}=-\frac{1}{a^2} \,d_2=\frac{1}{2\,x}\,d_3=\frac{1}{a^2} \,d_4 \,,\end{eqnarray}

Let us highlight that one can always get rid of second derivatives of the background fields by using their equations of motion. In this note, we have used the equations of motion for the background fields to express $G_{2},\,G_{4,X},\,G_{4,XX},\,G_{3,\phi},$ $G_{2,\phi X}$ (non-vanishing by assumption) in terms of other derivatives of the Lagrangian functions in order to obtain shorter expressions.

\subsubsection{Details for the speed of the graviton}

We wrote the dispersion relation of the gravitational waves as,
\begin{equation}
\omega^2 =\frac{f_0+\vec{p}\, {}^2 \,f_1}{f_2+\,\vec{p}\, {}^2\,f_3}\,\vec{p}\, {}^2\,, \nonumber
\end{equation}
and the speed $c_g^2=f_1/f_3$, where $f_0=\bar{f}_0/\bar{\mathcal{G}}_\tau,\, f_1=\bar{f}_1/\bar{\mathcal{G}}_\tau $, $f_2= -4\,c_1\,c_4 \,,\, f_3=d_1^2$ and $\bar{\mathcal{G}}_\tau = c_1\,(c_3^2-4\,b_1\,c_4)$. The latter are fixed by the coefficients $b_A,\, c_A,\,d_A$ given in the first part of this appendix, and

\begin{dmath}\bar{f}_{1}=- \frac{2}{\,{a}^{20}}\Big( \,{G_{5}}^2 \,{X} (-8 \,{a}^{10} \,{G_{4}} (\,{G_{4}}^2 + 2 \,{G_{4,X}} \,{G_{4}} \,{X} -  \,{G_{5}} (\,{G_{3}} + 2 \,{G_{5}} \,{H}^2) \,{X}) + 2 \,{a}^9 \,{G_{5}} \,{H} (4 \,{G_{4}}^2 + 8 \,{G_{4}} (\,{G_{4,X}} -  \,{G_{5,{\phi}}}) \,{X} + \,{G_{5}} \,{X} (\,{G_{3}} + \,{H}^2 (\,{G_{5}} + 2 \,{G_{5,X}} \,{X}))) \,\dot{\phi} + 2 \,{G_{5}}^2 \,{x}^2 \,\dot{\phi} (\,{x} \,{X} (3 \,{G_{5}} + 2 \,{G_{5,X}} \,{X}) -  \,{G_{5}} \,\ddot{\phi} \,\dot{\phi}) + \,{a}^3 \,{G_{5}}^2 \,{H} \,{x} \,\dot{\phi} (2 \,{x} \,{X} (5 \,{G_{5}} + 6 \,{G_{5,X}} \,{X}) -  \,{G_{5}} \,\ddot{\phi} \,\dot{\phi}) + 2 \,{a}^7 \,{G_{5}} \,\dot{\phi} (- \,{H} ((-4 \,{G_{4,X}} + \,{G_{5,{\phi}}}) \,{G_{5}} \,{X} + 2 \,{G_{4}} (\,{G_{5}} + 2 \,{G_{5,X}} \,{X})) \,\ddot{\phi} + \,{G_{4}} \,{G_{5}} \,\dot{H} \,\dot{\phi}) + 2 \,{a}^8 ((4 \,{G_{4}} \,{G_{5,{\phi}}} \,{G_{5}} \,{X} + \,{G_{3}} \,{G_{5}}^2 \,{X} - 2 \,{G_{4}}^2 (3 \,{G_{5}} + 2 \,{G_{5,X}} \,{X})) \,\ddot{\phi} + \,{G_{5}} \,{H}^2 (2 (\,{G_{4,X}} -  \,{G_{5,{\phi}}}) \,{G_{5}} \,{X} + \,{G_{4}} (\,{G_{5}} + 4 \,{G_{5,X}} \,{X})) \,\dot{\phi}^2) + 2 \,{a}^6 (-2 \,{G_{4}}^2 \,{x} (\,{G_{5}} + 2 \,{G_{5,X}} \,{X}) \,\dot{\phi} + \,{G_{4}} \,{G_{5}} (-2 \,{G_{5}} \,\ddot{\phi}^2 + 8 \,{G_{4,X}} \,{x} \,{X} \,\dot{\phi} - 4 \,{G_{5,{\phi}}} \,{x} \,{X} \,\dot{\phi} + \,{G_{5}} \,\dddot{\phi} \,\dot{\phi}) + \,{G_{5}}^2 (\,{G_{5,{\phi}}} \,{X} \,\ddot{\phi}^2 + 2 \,{G_{3}} \,{x} \,{X} \,\dot{\phi} + \,{H}^2 \,\dot{\phi} (\,{x} \,{X} (5 \,{G_{5}} + 6 \,{G_{5,X}} \,{X}) + \,{G_{5}} \,\ddot{\phi} \,\dot{\phi}))) + 2 \,{a}^4 \,{G_{5}} (\,{x} (6 \,{G_{4}} \,{G_{5}} \,{x} \,{X} + ((4 \,{G_{4,X}} + \,{G_{5,{\phi}}}) \,{G_{5}} \,{X} - 2 \,{G_{4}} (3 \,{G_{5}} + 2 \,{G_{5,X}} \,{X})) \,\ddot{\phi} \,\dot{\phi}) + 2 \,{G_{4}} \,{G_{5}} \,\dot{\phi}^2 \,\dot{x}) + \,{a}^2 \,{G_{5}}^2 \,\dot{\phi} (4 (3 \,{G_{4,X}} -  \,{G_{5,{\phi}}}) \,{x}^2 \,{X} \,\dot{\phi} + \,{G_{5}} \,\ddot{\phi} \,\dot{\phi} \,\dot{x} + \,{G_{5}} \,{x} (-3 \,\ddot{\phi}^2 + \,\dddot{\phi} \,\dot{\phi} + 4 \,{X} \,\dot{x})) + \,{a}^5 \,{G_{5}} \,\dot{\phi} (8 \,{G_{4}} \,{G_{5,X}} \,{H} \,{x} \,{X} \,\dot{\phi} + 2 \,{G_{5}} \,{H} \,{x} (6 \,{G_{4}} + 8 \,{G_{4,X}} \,{X} - 5 \,{G_{5,{\phi}}} \,{X}) \,\dot{\phi} + \,{G_{5}}^2 (-2 \,{H} \,\ddot{\phi}^2 + 2 \,{x} \,{X} \,\dot{H} + \,{H} \,\dddot{\phi} \,\dot{\phi} + \,\ddot{\phi} \,\dot{H} \,\dot{\phi} + 2 \,{H} \,{X} \,\dot{x})))\Big) \,.\end{dmath}

$\bar{f}_0$ is of little significance for the speed of the graviton and its form is cumbersome. Thus, we do not show it in this note. However, it can be easily obtained from the action (\ref{eqn Ltensor1}).

\subsection{Quadratic action for the scalar mode}\label{secapp scalar sector}

\subsubsection{Initial form of the action}

From (\ref{eqn action}), a direct computation gives the quadratic action for the scalar mode. It can be written in the unitary gauge (namely, with $\Pi=E=0$), as follows:
\begin{dmath}
\mathcal{S}_{\tau}=\frac{1}{2}\, \int\, \textrm{d}\eta\,\textrm{d}^3p \,\,\Big( M_{1}\, C^{\scalebox{0.5}{(3)}}\, \dot{\psi}\,+ M_{2}\, B\,\dot{\psi}\,+ M_{3}\, C^{\scalebox{0.5}{(2)}}\, C^{\scalebox{0.5}{(3)}}\,+ M_{4}\, B\, C^{\scalebox{0.5}{(3)}}\,+ M_{5}\, C^{\scalebox{0.5}{(2)}}\, \dot{\psi}\,+ M_{6}\, C^{\scalebox{0.5}{(2)}}\, \psi\,+ M_{7}\, B\, \psi\,+ M_{8}\, C^{\scalebox{0.5}{(3)}}\, \psi\,+ M_{9}\, \alpha\, \dot{\psi}\,+ M_{10}\, C^{\scalebox{0.5}{(3)}}\, \alpha\,+ M_{11}\, B\, \alpha\,+ M_{12}\, C^{\scalebox{0.5}{(2)}}\, \alpha\,+ M_{13}\, \alpha\, \psi\,+ M_{14}\, \dot{\psi}^2\,+ M_{15}\, \left(C^{\scalebox{0.5}{(3)}}\right)^2\,+ M_{16}\,\alpha^2\,+ M_{17}\,\psi^2\,\Big)\,,\label{eqn scalar action}
\end{dmath}
Eq. (\ref{eqn scalar action}) is trivially obtained after using the constraint equations for the Lagrange multipliers  $C^{\scalebox{0.5}{(1)}},\, C^{\scalebox{0.5}{(5)}},$ and $C^{\scalebox{0.5}{(7)}}$, which impose $C^{\scalebox{0.5}{(4)}} =C^{\scalebox{0.5}{(6)}}= C^{\scalebox{0.5}{(8)}}=0 $. The coefficients $M_A$ depend on conformal time and momentum, and can be written as:
\begin{dmath}M_{1}=\frac{\,{a}^4 (48 \,{G_{4,X}} - 24 \,{G_{5,{\phi}}}) \,{X} + \,{a}^3 (12 \,{G_{5}} \,{H} + 24 \,{G_{5,X}} \,{H} \,{X}) \,\dot{\phi} + (24 \,{G_{5}} \,{x} + 24 \,{G_{5,X}} \,{x} \,{X}) \,\dot{\phi}}{\,{a}^4} \,,\end{dmath}

\begin{dmath}M_{2}=-\frac{p^2\,a^2}{3}\,M_1+\frac{4 (2 \,{a}^4 \,{G_{4}} + \,{G_{5}} (\,{a}^3 \,{H} + \,{x}) \,\dot{\phi}) \, p^2}{\,{a}^2} \,,\end{dmath}

\begin{dmath}M_{3}=\frac{1}{a^4}\,M_2 - \frac{8 (2 \,{a}^4 \,{G_{4}} + \,{G_{5}} (\,{a}^3 \,{H} + \,{x}) \,\dot{\phi}) \, p^2}{\,{a}^6} \,,\end{dmath}

\begin{equation}
M_{4}= M_5=-\frac{p^2}{3}\,M_1 \,,
\end{equation}

\begin{dmath}M_{6}=-2\,x\,M_3 \,,\end{dmath}

\begin{equation}
M_{7}=\frac{2\,p^2\,x}{3}\,M_1 \,,
\end{equation}

\begin{dmath}p^2\,M_{8}=6\,x\,M_3+\frac{4}{a^2}\, p^4\,\,G_5\,\dot{\phi} \,,\end{dmath}

\begin{dmath}M_{9}=\frac{1}{\,{a}^4}\,\left(\,{a}^7 (-24 \,{G_{4}} \,{H} + (48 \,{G_{4,X}} - 48 \,{G_{5,{\phi}}}) \,{H} \,{X} + (96 \,{G_{4,XX}} - 48 \,{G_{5,{\phi}X}}) \,{H} \,{X}^2) + \,{a}^4 (-24 \,{G_{4}} \,{x} + (96 \,{G_{4,X}} \,{x} - 48 \,{G_{5,{\phi}}} \,{x}) \,{X} + (96 \,{G_{4,XX}} \,{x} - 48 \,{G_{5,{\phi}X}} \,{x}) \,{X}^2) + \,{a}^6 (-6 \,{G_{3}} - 6 \,{G_{5}} \,{H}^2 + (12 \,{G_{3,X}} - 24 \,{G_{4,{\phi}X}} + 48 \,{G_{5,X}} \,{H}^2) \,{X} + 24 \,{G_{5,XX}} \,{H}^2 \,{X}^2) \,\dot{\phi} + \,{a}^3 (12 \,{G_{5}} \,{H} \,{x} + 120 \,{G_{5,X}} \,{H} \,{x} \,{X} + 48 \,{G_{5,XX}} \,{H} \,{x} \,{X}^2) \,\dot{\phi} + (18 \,{G_{5}} \,{x}^2 + 72 \,{G_{5,X}} \,{x}^2 \,{X} + 24 \,{G_{5,XX}} \,{x}^2 \,{X}^2) \,\dot{\phi}\right) \,,\end{dmath}

\begin{dmath}M_{10}=\frac{1}{\,{a}^6}\,\left(\,{a}^7 ((96 \,{G_{4,X}} - 48 \,{G_{5,{\phi}}}) \,{H} \,{X} + (96 \,{G_{4,XX}} - 48 \,{G_{5,{\phi}X}}) \,{H} \,{X}^2) + \,{a}^4 ((144 \,{G_{4,X}} \,{x} - 48 \,{G_{5,{\phi}}} \,{x}) \,{X} + (96 \,{G_{4,XX}} \,{x} - 48 \,{G_{5,{\phi}X}} \,{x}) \,{X}^2) + \,{a}^6 (12 \,{G_{5}} \,{H}^2 + (12 \,{G_{3,X}} - 24 \,{G_{4,{\phi}X}} + 60 \,{G_{5,X}} \,{H}^2) \,{X} + 24 \,{G_{5,XX}} \,{H}^2 \,{X}^2) \,\dot{\phi} + \,{a}^3 (48 \,{G_{5}} \,{H} \,{x} + 144 \,{G_{5,X}} \,{H} \,{x} \,{X} + 48 \,{G_{5,XX}} \,{H} \,{x} \,{X}^2) \,\dot{\phi} + (36 \,{G_{5}} \,{x}^2 + 84 \,{G_{5,X}} \,{x}^2 \,{X} + 24 \,{G_{5,XX}} \,{x}^2 \,{X}^2) \,\dot{\phi}\right) \,,\end{dmath}

\begin{dmath}M_{11}=-\frac{1}{3}\,p^2\,M_9 \,,\end{dmath}

\begin{dmath}M_{12}=-\frac{1}{3}\,p^2\,M_{10} \,,\end{dmath}

\begin{dmath}M_{13}=\frac{1}{\,{a}^6}\,\left(\,{a}^7 (\,{H} (-192 \,{G_{4,X}} \,{x} + 96 \,{G_{5,{\phi}}} \,{x}) \,{X} + \,{H} (-192 \,{G_{4,XX}} \,{x} + 96 \,{G_{5,{\phi}X}} \,{x}) \,{X}^2) + \,{a}^4 ((-288 \,{G_{4,X}} \,{x}^2 + 96 \,{G_{5,{\phi}}} \,{x}^2) \,{X} + (-192 \,{G_{4,XX}} \,{x}^2 + 96 \,{G_{5,{\phi}X}} \,{x}^2) \,{X}^2) + \,{a}^6 (-24 \,{G_{5}} \,{H}^2 \,{x} + (-24 \,{G_{3,X}} \,{x} + 48 \,{G_{4,{\phi}X}} \,{x} - 120 \,{G_{5,X}} \,{H}^2 \,{x}) \,{X} - 48 \,{G_{5,XX}} \,{H}^2 \,{x} \,{X}^2) \,\dot{\phi} + \,{a}^3 (-96 \,{G_{5}} \,{H} \,{x}^2 - 288 \,{G_{5,X}} \,{H} \,{x}^2 \,{X} - 96 \,{G_{5,XX}} \,{H} \,{x}^2 \,{X}^2) \,\dot{\phi} + (-72 \,{G_{5}} \,{x}^3 - 168 \,{G_{5,X}} \,{x}^3 \,{X} - 48 \,{G_{5,XX}} \,{x}^3 \,{X}^2) \,\dot{\phi} + (\,{a}^8 (-8 \,{G_{4}} + (16 \,{G_{4,X}} - 8 \,{G_{5,{\phi}}}) \,{X}) + 8 \,{a}^7 \,{G_{5,X}} \,{H} \,{X} \,\dot{\phi} + \,{a}^4 (4 \,{G_{5}} \,{x} + 8 \,{G_{5,X}} \,{x} \,{X}) \,\dot{\phi}) \, p^2\right) \,,\end{dmath}

\begin{dmath}
p^2\,M_{14}=-\frac{3}{2}\,M_2 \,,
\end{dmath}

\begin{dmath}
p^2\,M_{15}=-\frac{3}{2}\,M_3 \,,
\end{dmath}

\begin{dmath}M_{16}=\frac{1}{\,{a}^6}\,\Big(\,{a}^{10} (-12 \,{G_{4}} \,{H}^2 + (-2 \,{G_{2,X}} - 36 \,{G_{5,{\phi}}} \,{H}^2) \,{X} + (4 \,{G_{2,XX}} - 4 \,{G_{3,{\phi}X}} + (144 \,{G_{4,XX}} - 84 \,{G_{5,{\phi}X}}) \,{H}^2) \,{X}^2 + (48 \,{G_{4,XXX}} - 24 \,{G_{5,{\phi}XX}}) \,{H}^2 \,{X}^3) + \,{a}^3 (46 \,{G_{5}} \,{H} \,{x}^2 + 208 \,{G_{5,X}} \,{H} \,{x}^2 \,{X} + 156 \,{G_{5,XX}} \,{H} \,{x}^2 \,{X}^2 + 24 \,{G_{5,XXX}} \,{H} \,{x}^2 \,{X}^3) \,\dot{\phi} + (18 \,{G_{5}} \,{x}^3 + 82 \,{G_{5,X}} \,{x}^3 \,{X} + 56 \,{G_{5,XX}} \,{x}^3 \,{X}^2 + 8 \,{G_{5,XXX}} \,{x}^3 \,{X}^3) \,\dot{\phi} + \,{a}^9 (-8 \,{G_{4}} \,\dot{H} + (-2 \,{G_{3}} \,{H} - 2 \,{G_{5}} \,{H}^3 + ((12 \,{G_{3,X}} - 36 \,{G_{4,{\phi}X}}) \,{H} + 44 \,{G_{5,X}} \,{H}^3) \,{X} + ((12 \,{G_{3,XX}} - 24 \,{G_{4,{\phi}XX}}) \,{H} + 44 \,{G_{5,XX}} \,{H}^3) \,{X}^2 + 8 \,{G_{5,XXX}} \,{H}^3 \,{X}^3) \,\dot{\phi}) + \,{a}^7 (8 \,{G_{4}} \,{H} \,{x} - 72 \,{G_{5,{\phi}}} \,{H} \,{x} \,{X} + \,{H} (336 \,{G_{4,XX}} \,{x} - 168 \,{G_{5,{\phi}X}} \,{x}) \,{X}^2 + \,{H} (96 \,{G_{4,XXX}} \,{x} - 48 \,{G_{5,{\phi}XX}} \,{x}) \,{X}^3 - 8 \,{G_{4,X}} \,{H} \,\ddot{\phi} \,\dot{\phi}) + \,{a}^4 (-4 \,{G_{4}} \,{x}^2 - 36 \,{G_{5,{\phi}}} \,{x}^2 \,{X} + (192 \,{G_{4,XX}} \,{x}^2 - 84 \,{G_{5,{\phi}X}} \,{x}^2) \,{X}^2 + (48 \,{G_{4,XXX}} \,{x}^2 - 24 \,{G_{5,{\phi}XX}} \,{x}^2) \,{X}^3 - 8 \,{G_{4,X}} \,{x} \,\ddot{\phi} \,\dot{\phi}) + \,{a}^8 (-2 \,{G_{3}} \,\ddot{\phi} - 2 \,{G_{5}} \,{H}^2 \,\ddot{\phi} - 4 \,{G_{5,X}} \,{H}^2 \,{X} \,\ddot{\phi} - 4 \,{G_{5}} \,{H} \,\dot{H} \,\dot{\phi} + 22 \,{G_{4,X}} \,{H}^2 \,\dot{\phi}^2) + \,{a}^6 ((-4 \,{G_{3}} \,{x} + 26 \,{G_{5}} \,{H}^2 \,{x} + (18 \,{G_{3,X}} \,{x} - 36 \,{G_{4,{\phi}X}} \,{x} + 170 \,{G_{5,X}} \,{H}^2 \,{x}) \,{X} + (12 \,{G_{3,XX}} \,{x} - 24 \,{G_{4,{\phi}XX}} \,{x} + 144 \,{G_{5,XX}} \,{H}^2 \,{x}) \,{X}^2 + 24 \,{G_{5,XXX}} \,{H}^2 \,{x} \,{X}^3) \,\dot{\phi} - 8 \,{G_{4}} \,\dot{x}) + \,{a}^2 (-2 \,{G_{5}} \,{x}^2 \,\ddot{\phi} - 4 \,{G_{5,X}} \,{x}^2 \,{X} \,\ddot{\phi} + 50 \,{G_{4,X}} \,{x}^2 \,\dot{\phi}^2 - 4 \,{G_{5}} \,{x} \,\dot{\phi} \,\dot{x}) + \,{a}^5 (-4 \,{G_{5}} \,{H} \,{x} \,\ddot{\phi} - 8 \,{G_{5,X}} \,{H} \,{x} \,{X} \,\ddot{\phi} + 72 \,{G_{4,X}} \,{H} \,{x} \,\dot{\phi}^2 + \,\dot{\phi} (-4 \,{G_{5}} \,{x} \,\dot{H} - 4 \,{G_{5}} \,{H} \,\dot{x}))\Big) \,,\end{dmath}

\begin{dmath}M_{17}=\frac{1}{\,{a}^6}\,\left((72 \,{G_{5}} \,{x}^3 + 48 \,{G_{5,X}} \,{x}^3 \,{X}) \,\dot{\phi} + \,{a}^6 (-12 \,{G_{5}} \,{H}^2 \,{x} + (48 \,{G_{4,{\phi}X}} \,{x} - 24 \,{G_{5,{\phi}{\phi}}} \,{x} - 96 \,{G_{5,X}} \,{H}^2 \,{x}) \,{X} - 48 \,{G_{5,XX}} \,{H}^2 \,{x} \,{X}^2) \,\dot{\phi} + \,{a}^3 (-48 \,{G_{5}} \,{H} \,{x}^2 - 144 \,{G_{5,X}} \,{H} \,{x}^2 \,{X} - 48 \,{G_{5,XX}} \,{H} \,{x}^2 \,{X}^2) \,\dot{\phi} + \,{a}^4 (48 \,{G_{4}} \,{x}^2 + (48 \,{G_{4,X}} \,{x} \,\ddot{\phi} - 24 \,{G_{5,{\phi}}} \,{x} \,\ddot{\phi} + \,{X} (48 \,{G_{4,XX}} \,{x} \,\ddot{\phi} - 24 \,{G_{5,{\phi}X}} \,{x} \,\ddot{\phi})) \,\dot{\phi} + \,\dot{\phi}^2 (24 \,{G_{4,X}} \,\dot{x} - 12 \,{G_{5,{\phi}}} \,\dot{x})) + \,{a}^2 (24 \,{G_{5}} \,{x}^2 \,\ddot{\phi} + 120 \,{G_{5,X}} \,{x}^2 \,{X} \,\ddot{\phi} + 48 \,{G_{5,XX}} \,{x}^2 \,{X}^2 \,\ddot{\phi} + (48 \,{G_{4,X}} \,{x}^2 + 24 \,{G_{5,{\phi}X}} \,{x}^2 \,{X}) \,\dot{\phi}^2 + \,\dot{\phi} (48 \,{G_{5}} \,{x} \,\dot{x} + 48 \,{G_{5,X}} \,{x} \,{X} \,\dot{x})) + \,{a}^5 (12 \,{G_{5}} \,{H} \,{x} \,\ddot{\phi} + 96 \,{G_{5,X}} \,{H} \,{x} \,{X} \,\ddot{\phi} + 48 \,{G_{5,XX}} \,{H} \,{x} \,{X}^2 \,\ddot{\phi} + (\,{H} (-48 \,{G_{4,X}} \,{x} + 36 \,{G_{5,{\phi}}} \,{x}) + \,{H} (-48 \,{G_{4,XX}} \,{x} + 48 \,{G_{5,{\phi}X}} \,{x}) \,{X}) \,\dot{\phi}^2 + \,\dot{\phi} (12 \,{G_{5}} \,{x} \,\dot{H} + 12 \,{G_{5}} \,{H} \,\dot{x} + \,{X} (24 \,{G_{5,X}} \,{x} \,\dot{H} + 24 \,{G_{5,X}} \,{H} \,\dot{x}))) + (\,{a}^8 (4 \,{G_{4}} - 4 \,{G_{5,{\phi}}} \,{X}) - 4 \,{a}^6 \,{G_{5,X}} \,{X} \,\ddot{\phi} - 6 \,{a}^4 \,{G_{5}} \,{x} \,\dot{\phi} + 4 \,{a}^7 \,{G_{5,X}} \,{H} \,{X} \,\dot{\phi}) \, p^2\right) \,.\end{dmath}

Again, as in the Appendix \ref{secapp coefficients tensor modes},  we use the equations of motion for the background fields to express $G_{2},\,G_{4,X},\,G_{4,XX},\,G_{3,\phi},$ $G_{2,\phi X}$ (non-vanishing by assumption) in terms of other derivatives of the Lagrangian functions in order to obtain shorter expressions.

\subsubsection{Obtaining the quadratic action for the scalar mode in its final form}

Now, from the action (\ref{eqn scalar action}) we can obtain the form (\ref{eqn Lscalar}) as follows: the Euler-Langrange equation for the Torsion scalar $C^{\scalebox{0.5}{(2)}}$ -which is a Lagrange multiplier- is a constraint that can be used to express  $C^{\scalebox{0.5}{(3)}}$ in terms of $\alpha,\,\psi,\, \dot{\psi}$. Plugging back $C^{\scalebox{0.5}{(3)}}(\alpha,\,\psi,\, \dot{\psi})$ in (\ref{eqn scalar action}) gives (\ref{eqn Lscalar}). In particular, as explained before, the functions of time $\bar{\mathcal{G}}_{\mathcal{S}}$ and $T$ in Eq. (\ref{eqn Lscalar}) are essential to avoid the No-Go theorems. Let us explicitly write them:

\begin{dmath}\bar{\mathcal{G}}_{\mathcal{S}}=\frac{(2 \,{a}^4 \,{G_{4}} + \,{G_{5}} (\,{a}^3 \,{H} + \,{x}) \,\dot{\phi})^2}{2 \,{a}^8 \,{G_{4}} + 2 \,{a}^7 \,{H} (\,{G_{5}} + \,{G_{5,X}} \,{X}) \,\dot{\phi} + \,{a}^4 \,{x} (3 \,{G_{5}} + 2 \,{G_{5,X}} \,{X}) \,\dot{\phi} + \,{a}^6 (2 \,{G_{4,X}} -  \,{G_{5,{\phi}}}) \,\dot{\phi}^2} \,,\end{dmath}
\begin{equation}
T=\frac{t_1}{t_2}\,,\label{eqn TG5}
\end{equation}
with
\begin{dmath}t_1=\Big(2 (\,{x}^2 \,{X}^2 (\,{G_{5,X}} \,{G_{5}} + 4 \,{G_{5,X}}^2 \,{X} - 2 \,{G_{5,XX}} \,{G_{5}} \,{X}) - 2 \,{a}^3 \,{H} \,{x} \,{X} (\,{G_{5}}^2 - 4 \,{G_{5,X}}^2 \,{X}^2 + 2 \,{G_{5,XX}} \,{G_{5}} \,{X}^2) + \,{a}^6 (-2 \,{G_{4}}^2 + \,{X} (- \,{G_{5}}^2 \,{H}^2 + 2 \,{X} ((-2 \,{G_{4,X}} + \,{G_{5,{\phi}}})^2 + 2 \,{G_{5,X}}^2 \,{H}^2 \,{X}) -  \,{G_{5}} \,{X} (\,{G_{3,X}} - 2 \,{G_{4,{\phi}X}} + \,{H}^2 (\,{G_{5,X}} + 2 \,{G_{5,XX}} \,{X})))) - 2 \,{a}^5 \,{H} (\,{G_{4}} \,{G_{5}} + (-4 \,{G_{4,X}} \,{G_{5,X}} + 2 \,{G_{5,{\phi}}} \,{G_{5,X}} + 2 \,{G_{4,XX}} \,{G_{5}} -  \,{G_{5,{\phi}X}} \,{G_{5}}) \,{X}^2) \,\dot{\phi} - 2 \,{a}^2 \,{x} (\,{G_{4}} \,{G_{5}} + (- \,{G_{4,X}} + \,{G_{5,{\phi}}}) \,{G_{5}} \,{X} + (-4 \,{G_{4,X}} \,{G_{5,X}} + 2 \,{G_{5,{\phi}}} \,{G_{5,X}} + 2 \,{G_{4,XX}} \,{G_{5}} -  \,{G_{5,{\phi}X}} \,{G_{5}}) \,{X}^2) \,\dot{\phi})\Big) \,,\end{dmath}

\begin{dmath}
t_2=\Big(2 \,{a}^6 (\,{G_{4}} + 2 \,{G_{4,X}} \,{X} -  \,{G_{5,{\phi}}} \,{X}) + 2 \,{a}^5 \,{H} (\,{G_{5}} + \,{G_{5,X}} \,{X}) \,\dot{\phi} + \,{a}^2 \,{x} (3 \,{G_{5}} + 2 \,{G_{5,X}} \,{X}) \,\dot{\phi}\Big)\,.
\end{dmath}

Thus, by inspection it is clear that even if Eq. (\ref{eqn linkTtoSlowP}) obligues us to satisfy $\bar{\mathcal{G}}_{\mathcal{S}}>0$  in order to have a stable graviton at low momentum, $t_1$ and hence $T$ can still vanish in many ways with a careful choice of Lagrangian functions (it is worth to note at this point that in general $T\neq \Theta$). 

That we can choose Lagrangian functions that allow $T$ to vanish at some point -while still satisfying the stability of all modes at all momenta, and subluminality- is the key tool to build theories without suffering the usual No-Go theorems.

The technical detail that leads to these new possibilities boils down to a new $G_5$ term in the action (\ref{eqn scalar action}) in comparison to up to quartic Horndeski-Cartan theories: namely, the $p^2$ part in the term $M_{8}\, C^{\scalebox{0.5}{(3)}}\, \psi\,$. Indeed, the constraint imposed by $C^{\scalebox{0.5}{(2)}}$ is of the form  $C^{\scalebox{0.5}{(3)}}\,=\, -\frac{M_{12}}{M_3}\, \alpha+\dots$, which then gives a new $\mathcal{O}(p^2\,G_5)$ term in $-\frac{M_{12}}{M_3}\,M_8\,\alpha\,\psi$ which gives a totally new $T$ in comparison to simpler forms of the theory.

\section{References}

\bibliographystyle{IEEEtran}
\bibliography{v4HorndeskiCCosmoPert}


\end{document}